# A Bayesian Network Method for Deaggregation: Identification of Tropical Cyclones Driving Coastal Hazards

## Preprinted Manuscript

Ziyue Liu, Ph.D.,[1] Meredith L. Carr, Ph.D., P.E.,[2] Norberto C. Nadal-Caraballo, Ph.D.,[3] Madison C. Yawn,[4] and Michelle T. Bensi, Ph.D.[5]

[1]Post-doctoral Researcher, Department of Civil and Environmental Engineering, University of Maryland; email: ziyue20@terpmail.umd.edu; ORCID: 0009-0003-5850-0374

[2]Research Hydraulic Engineer, Coastal & Hydraulics Laboratory, U.S. Army Engineer R&D Center; email: Meredith.L.Carr@erdc.dren.mil

[3]Senior Research Civil Engineer, Coastal & Hydraulics Laboratory, U.S. Army Engineer R&D Center; email: Norberto.C.Nadal-Caraballo@usace.army.mil

[4]Research Physical Scientist, Coastal & Hydraulics Laboratory, U.S. Army Engineer R&D Center; email: Madison.C.Yawn@usace.army.mil

[5]Associate Professor, Department of Civil and Environmental Engineering, University of Maryland; email: mbensi@umd.edu; ORCID: 0000-0001-6449-1812

### Highlights

- Adapts deaggregation approaches used for seismic hazards to the tropical cyclone-induced hazards context.
- Novel method using Bayesian networks to identify tropical cyclone scenarios driving severe coastal hazards and reveal dependence among coastal hazard mechanisms.
- Supports risk-informed decision-making regarding tropical cyclone-induced coastal hazards and guides targeted high-fidelity modeling.

### Abstract

The Joint Probability Method (JPM) has been the primary approach for developing hazard curves for tropical cyclone (TC)-induced coastal hazards for decades. To develop hazard curves that capture the breadth of TC-induced coastal hazards, a large number of synthetic TCs must be simulated, which is computationally intensive and time-consuming. This problem is exacerbated when considering compound coastal hazards. Given that low-exceedance-probability (LEP) coastal hazards are likely to result in the most significant damage to coastal communities, it is practical to focus resource-intensive modeling efforts on TC scenarios that are dominant contributors to LEP coastal hazards. In this study, we adapt deaggregation methods first established in probabilistic seismic hazard analysis to the context of TC hazard assessment. To enable deaggregation, we develop a Bayesian network (BN)-based framework incorporating the existing JPM for multiple TC-induced coastal hazards. BNs are selected as the quantification mechanism because they enable visualization of causal relationships and efficient, multi-directional probabilistic inference. The primary objective of this study is to leverage BNs to better understand the dominant drivers of individual and compound coastal hazards at a given location. Case studies are applied to the Greater New Orleans region in Louisiana (USA). Our findings highlight the potential of employing BN inference analysis to (1) identify TCs that are dominant contributors to multiple LEP coastal hazards and (2) understand the dependence between TC-induced coastal hazard mechanisms.

### Plain Language Summary

Tropical cyclones cause significant damage to coastal communities worldwide. Although substantial efforts have been made to assess hazards caused by tropical cyclones (e.g., coastal water levels or rainfall depths), accurately evaluating their risks can be expensive due to the high computational demands and the wide range of potential storm scenarios. In this work, we develop a method that uses a Bayesian network to identify the most likely tropical cyclone scenarios that lead to severe impacts on coastal regions. The results of this approach can help ensure resource-intensive modeling efforts focus on critical tropical cyclone scenarios, optimizing resource use while improving hazard assessments.

## 1. Introduction

Tropical cyclones (TCs) cause significant damage to coastal communities due to flooding from storm surge, rainfall, and other drivers occurring individually or in combination. Systematic risk-informed planning and mitigation decisions involving coastal infrastructure, insurance pricing, and other policy decisions require an understanding of the likelihood of flooding in coastal regions.

State-of-the-art probabilistic coastal hazard assessments (PCHA) used to estimate the frequency of coastal hazards typically leverage the Joint Probability Method (JPM). The JPM provides a quantitative framework to combine the results of statistical assessments of storm climatology with physical representations of coastal processes. Most applications of the JPM to date have focused on a single measure of hazard severity (e.g., storm surge elevation). However, efforts are underway to expand this scope to multiple or compounding hazards.

The JPM systematically accounts for all possible hurricane scenarios that can contribute to coastal hazards and their relative likelihoods of occurrence. It typically involves direct numerical integration to estimate the frequency at which a particular coastal location will experience coastal hazards over a range of severities. In this work, we propose implementing the JPM for multi-hazard assessment using Bayesian networks (BNs). BNs are probabilistic graphical models that can be used to replicate the results of a conventional JPM analysis performed using direct numerical integration and offer other advantages that are not readily achievable using the conventional approach. These advantages include transparent representation of complex dependencies and the ability to perform multi-directional probabilistic inference. In this study, we demonstrate how the backward inference capabilities of BNs can be used to expand the capabilities of PCHA to identify dominant sets of storm parameters that lead to severe local hazards. These results can be used to better understand the dominant drivers of coastal hazards at a given location or to refine storm parameter set selection to more comprehensively represent multiple forcings, such as compound coastal hazards analysis of surge and rainfall. Building on terminology commonly used in seismic hazard applications (Baker 2013; Bazzurro and Allin Cornell 1999; McGuire 1995), this process of identifying dominant contributors is referred to in this study as deaggregation.

Section 2 of this paper summarizes key background information related to BNs. Section 3 describes how consideration of the fundamental JPM integral yields the BN structure for single- and multiple-hazard cases. Section 4 describes the implementation of the BN for a case study by defining model choices and assumptions necessary to enable quantitative inference in the BN. Section 5 describes the use of the BN for deaggregation, and Section 6 provides a summary and conclusions.

## 2. Background on Bayesian networks

A BN is a Bayesian rule-motivated directed acyclic graphical model representing probabilistic relationships among multiple dependent variables. BNs comprise nodes (circles) representing random variables and links (arcs) representing dependencies, typically reflecting causal relationships when used in an engineering context. A simple example of a BN with four nodes (random variables) is shown in Fig. 1 to introduce key concepts. In this BN, the nodes $X_2$ and $X_3$ are shown as being conditionally dependent on a common causal variable $X_1$ as represented by the links going from the node $X_1$ to the nodes $X_2$ and $X_3$. BNs use a "familial" terminology to describe the relationship between nodes. Applying this terminology, $X_1$ is a *parent node* of $X_2$ and $X_3$. In turn, $X_2$ and $X_3$ are described as *child nodes* of $X_1$. Similarly, $X_4$ is a child of nodes $X_2$ and $X_3$ representing the conditional dependence of $X_4$ on both of its parents $X_2$ and $X_3$. Each node in the BN is associated with a set of mutually exclusive states representing possible outcomes. If the random variables modeled in the BN are continuous, they will generally require discretization such that a node's states reflect intervals within the random variable's continuous domain.

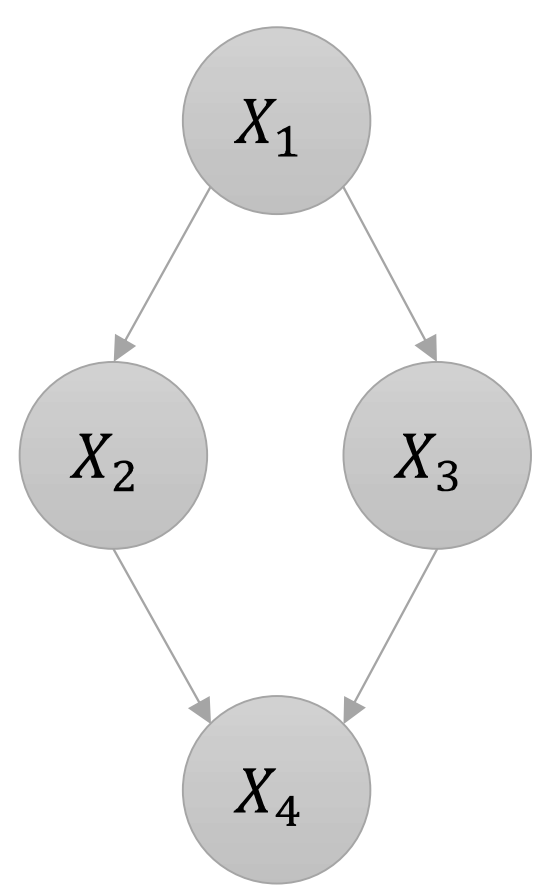

**Fig. 1 Example of a simple BN**

The joint probability distribution over all nodes (random variables), $\mathbf{X} = \{X_1, \dots, X_n\}$, within a BN can be factorized by making use of local conditional relationships:

$$p(\mathbf{x}) = \prod_{i=1}^{n} p\big(x_i \big| Pa(X_i)\big) \tag{1}$$

where $p(\mathbf{x})$ is the joint distribution over all nodes in the BN and $p\big(x_i \big| Pa(X_i)\big)$ is the conditional probability distribution of node $X_i$ given its set of parent nodes, $Pa(X_i)$. Thus, for the BN in Fig. 1, the joint distribution over all nodes can be written as $p(x_1, x_2, x_3, x_4) = p(x_4|x_3, x_2)p(x_3|x_1)p(x_2|x_1)p(x_1)$. In a BN, the conditional probability distribution $p\big(x_i \big| Pa(X_i)\big)$ is represented using a conditional probability table (CPT) that provides the conditional probability that a node $X_i$ takes on each of its possible states given each possible combination of states of its parent nodes. If a node $X_i$ represents a continuous random variable with states corresponding to intervals within the random variable's continuous domain, a node taking on a given state $j$ reflects the probability that the random variable is within a particular range given the states of its parents, i.e., $P\left(x_{i,j}^{-} < X_i < x_{i,j}^{+} \middle| pa(X_i)\right)$, where $x_{i,j}^{-}$ and $x_{i,j}^{+}$ are the lower boundary and upper boundary of the particular interval.

Efficient algorithms are available to enable marginalization in the BN (i.e., the determination of the joint distribution of any subset of variables in the BN). BNs can also be used for multiple types of probabilistic inference under "evidence cases." The term "evidence case" is used to refer to a situation in which the states of some variables in the BN are known or assumed to be known. Forward (predictive) inference calculates the probability distributions of nodes by propagating the evidence in the direction of the arrows. For example, considering the BN in Fig. 1, an example of forward inference would involve the calculation of the probability that $X_4$ takes on any of its states, given that $X_1$'s state is known. Many computational techniques can enable forward inference. A unique advantage of the BN is the capability to efficiently perform backward inference, which calculates the probability distributions of nodes by propagating evidence in the opposite direction of the arrows. For example, considering the BN in Fig. 1, an example of backward inference would involve the calculation of the probability that $X_1$ takes on any of its states, given that $X_4$'s state is known.

In this study, BNs are used for probabilistic modeling and inference. We establish the BN structure and CPTs utilizing a mix of physical processes and statistical knowledge. Then, the BN is used to perform forward and backward inference tasks. It is noted that BNs can also be used in statistical learning contexts if sufficient data is available. For example, BN parameter learning involves specifying the network structure and using algorithms to estimate the CPTs directly from data. BN structure learning, on the other hand, focuses on learning the network structure itself from data. BNs can also be expanded with decision and utility nodes to enable the modeling of decision problems.

## 3. Structure of BN for multi-hazard JPM analysis

This study aims to implement a BN to expand the capabilities of JPM-based PCHA to enable deaggregation (i.e., identification of dominant sets of storm parameters that lead to severe local hazards) through backward inference in the BN. This study focuses on using BNs for deaggregation in the multi-hazard context. The foundational concept of modeling TC-induced coastal hazards considering multiple flood-forcing processes via a BN was introduced by Mohammadi et al. (2023). We leverage and expand upon the probabilistic modeling and inference approach of Mohammadi et al. (2023). This study also provides quantitative implementation details and commentary associated BN-based deaggregation method as applied in (Liu et al. 2025).

In recent years, BNs have also been implemented in various other contexts for coastal hazard analysis, including parameter/structure learning and decision models. For example, Sebastian et al. (2017) described a non-parametric BN-based storm events simulation method to determine the hydraulic boundary conditions for a low-lying coastal watershed in the Houston-Galveston region of Texas. Couasnon et al. (2018) built a BN based on a Gaussian copula to generate coastal watershed boundary conditions in Southeast Texas. Flores et al. (2019) created a hybrid dynamic object-oriented BN to model the risk of flooding in a Mediterranean catchment in the south of Spain. Sanuy et al. (2020) built a BN to characterize the relationship between wave conditions, coastal erosion, and coastal inundation in the Catalan coast region of Spain. Callens et al. (2023) incorporated statistical learning method-generated data for BN construction and found that the BN supported by statistical learning method data outperformed the BN built exclusively on observational data in predicting coastal floods. Salgado et al. (2024) leveraged BN to explore the relative importance of different storm-related risk factors in Mexico. Durap (2024) incorporated BN and geospatial information for coastal hazard identification and decision-making in Queensland, Australia. Garzon et al. (2024) combined BN and numerical model simulation data to develop a coastal erosion early warning system.

In the sections that follow, we define how the JPM leads to the structure of the BN. We begin by briefly summarizing the JPM (Section 3.1) before describing the BN-based implementation of the JPM for individual hazards (Section 3.2) and multi-hazards (Section 3.3). Then, in Section 4, we discuss the implementation of the BN by outlining strategies for developing the CPTs.

### 3.1. Overview of the Joint Probability Method

The JPM has been implemented for coastal hazard analysis for decades (Myers 1954; Ho and Myers 1975; Russell 1969; Myers 1970), with most implementations focused primarily on storm surge hazards (Nadal-Caraballo et al. 2015, 2020; Toro 2008; Toro et al. 2010).

The foundational JPM integral equation (Nadal-Caraballo et al. (2020)) takes the following form:

$$\lambda_{R>r^*} = \lambda \int P(R > r^* | \mathbf{x}, \varepsilon)\, f_{\mathbf{X}}(\mathbf{x}) f_{\varepsilon}(\varepsilon) d\mathbf{x} \mathrm{d}\varepsilon \tag{2}$$

Where:

- $\lambda_{R>r^*}$ is the annual frequency of exceedance of TC response, which represents how often a measure of storm response $R$ (e.g., storm surge elevation) exceeds $r^*$ at a given location;
- $\lambda$ is the storm recurrence rate, which represents how often TCs affect a given location;
- $f_{\mathbf{X}}(\mathbf{x})$ is the joint distribution of TC parameters, which represents the likelihood of a TC having characteristics defined by the parameter vector **x**.
- **x** is a vector of TC parameter values, defined in this study as consisting of central pressure deficit ($\Delta P$), forward velocity ($V_{\mathrm{f}}$), radius of maximum winds ($R_{\max}$), heading ($\Theta$), and landfall location ($X_0$).
- $P(R > r^* | \mathbf{x}, \varepsilon)$ is the conditional probability of $R > r^*$ given TC parameter forcing vector **x** and error $\varepsilon$, which represents the probability that the TC response, defined as the superposition of a model prediction and model error (i.e., $\hat{r}(\mathbf{x}) + \varepsilon$), will exceed a particular value $r^*$ given a TC having characteristics defined by the parameter vector **x** and considering a particular value of the model prediction error;
- $f_{\varepsilon}(\varepsilon)$ is the probability distribution of model error.

The JPM integral is typically calculated in discrete form using numerical integration. In this study, we express the discrete form of the JPM integral as:

$$\lambda_{R>r^*} \approx (\lambda * S_{trk}) \sum_{j=1}^{n_{\varepsilon}} \sum_{i=1}^{n_{\mathbf{x}}} P\left[\hat{r}(\mathbf{x}_i) + \varepsilon_j > r^* | \mathbf{x}_i, \varepsilon_j\right] p(\mathbf{x}_i) p(\varepsilon_j) \tag{3}$$

In the above expression, the domains of each of the parameters in $\mathbf{x} = \{\delta p, v_{\mathrm{f}}, r_{\max}, \theta, x_0\}$ are discretized, respectively, into $n_{\Delta P}$, $n_{V_{\mathrm{f}}}$, $n_{R_{\max}}$, $n_{\Theta}$, and $n_{X_o}$bins, each representing a small portion of the continuous parameter domain. This yields $n_{\mathbf{x}} = n_{\Delta P} \times n_{V_{\mathrm{f}}} \times n_{R_{\max}} \times n_{\Theta} \times n_{X_0}$ discrete combinations of TC parameter bins. In the expression in Equation (2), $\mathbf{x}_i$ represents one discrete combination of TC parameter bins, which can be used to define a synthetic TC consisting of a track (based on landfall location and heading) along with intensity, size, and speed information. Each synthetic TC is associated with a discrete probability $p(\mathbf{x}_i)$, representing a portion of the continuous joint distribution $f_{\mathbf{X}}(\mathbf{x})$. The domain of the model error term is likewise discretized into a set of bins, each of which is associated with a discrete probability mass $p(\varepsilon_j)$. $P[\hat{r}(\mathbf{x}_i) + \varepsilon_j > r^* | \mathbf{x}_i, \varepsilon_j]$ represents the conditional probability that $\hat{r}(\mathbf{x}_i) + \varepsilon_j$ is greater than $r^*$, given TC parameter combination (TCPC) $\mathbf{x}_i$ and error $\varepsilon_j$. $\lambda$ is the storm recurrence rate (in units of storms/year/unit distance) and represents how often TCs affect a location. $S_{trk}$ represents TC track spacing and is used to assign a portion of the storm recurrence rate attributable to each synthetic TC track.

## 3.2. Structure of BN reflecting single-hazard JPM analysis

Using the factorization approach described in Section 2, a BN can be constructed to reflect the joint distribution over random variables involved in the JPM integral described in Section 3.1. First, the joint distribution of TC parameters, $f_{\mathbf{X}}(\mathbf{x})$, must be specified. Consistent with recent practice ( Nadal-Caraballo et al. 2022b; Liu et al. 2024b), an assumption is made that the parameters $\{\Delta P, R_{\max}, V_{\mathrm{f}}, \theta\}$ are fully dependent and independent of landfall location. The joint distribution over the TC parameters is then factorized as:

$p(\delta p, v_{\mathrm{f}}, r_{\max}, \theta, x_0) = p(\delta p, v_{\mathrm{f}}, r_{\max}, \theta) * p(x_0) = p(\theta | \delta p, v_{\mathrm{f}}, r_{\max}) p(r_{\max} | \delta p, v_{\mathrm{f}}) p(v_{\mathrm{f}} | \delta p) p(\delta p) p(x_0)$

The resulting BN is shown in Fig. 2, with physical random variables as defined in Table 1.

**Table 1: Physical definition of each node in BN**

| Notation | Physical definition | Notation | Physical definition |
|---|---|---|---|
| $\Delta P$: | Storm central pressure deficit | $X_0$: | Storm landfall location |
| $V_{\mathrm{f}}$: | Storm forward velocity | $\hat{R}$: | Coastal hazard severity measure, as predicted by a model |
| $R_{\max}$: | Storm radius of maximum winds | $R$: | Coastal hazard severity measure |
| $\Theta$: | Storm heading direction | $\varepsilon_R$: | Error node accounting for the prediction model error |

The BN has a joint distribution specified as:

$$p(\boldsymbol{\omega}) = p(r | \hat{r}, \varepsilon_R) \times p(\varepsilon_R) \times p(\hat{r} | \delta p, v_{\mathrm{f}}, r_{\max}, \theta, x_0) \times p(\theta | \delta p, v_{\mathrm{f}}, r_{\max}) \\ \times p(r_{\max} | \delta p, v_{\mathrm{f}}) \times p(v_{\mathrm{f}} | \delta p) p(\delta p) \times p(x_0) \tag{4}$$

where lowercase symbols are used to represent realizations of the random variable; $\boldsymbol{\omega}$ represents the set of all variables in the BN.

Following marginalization, the node $R$ will output the probability of each discretized bin of coastal hazard as $p(r_k)$, where $r_k$ is the representative value (usually set as the midpoint) of the $k$th discretized bin. The portion of the JPM integral in Equation (2) can be approximately computed as the sum of the probabilities of all $p(r_k)$ with $r_k > r^*$:

$$\int P(R > r^* | \mathbf{x}, \varepsilon)\, f_{\mathbf{X}}(\mathbf{x}) f_{\varepsilon}(\varepsilon) d\mathbf{x} \mathrm{d}\varepsilon \approx \sum_{r_k > r^*} p(r_k) \tag{5}$$

This quantity can be multiplied by the storm recurrence rate and storm track spacing ($\lambda * S_{trk}$) to yield the annual frequency of exceedance of TC response ($\lambda_{R>r}$).

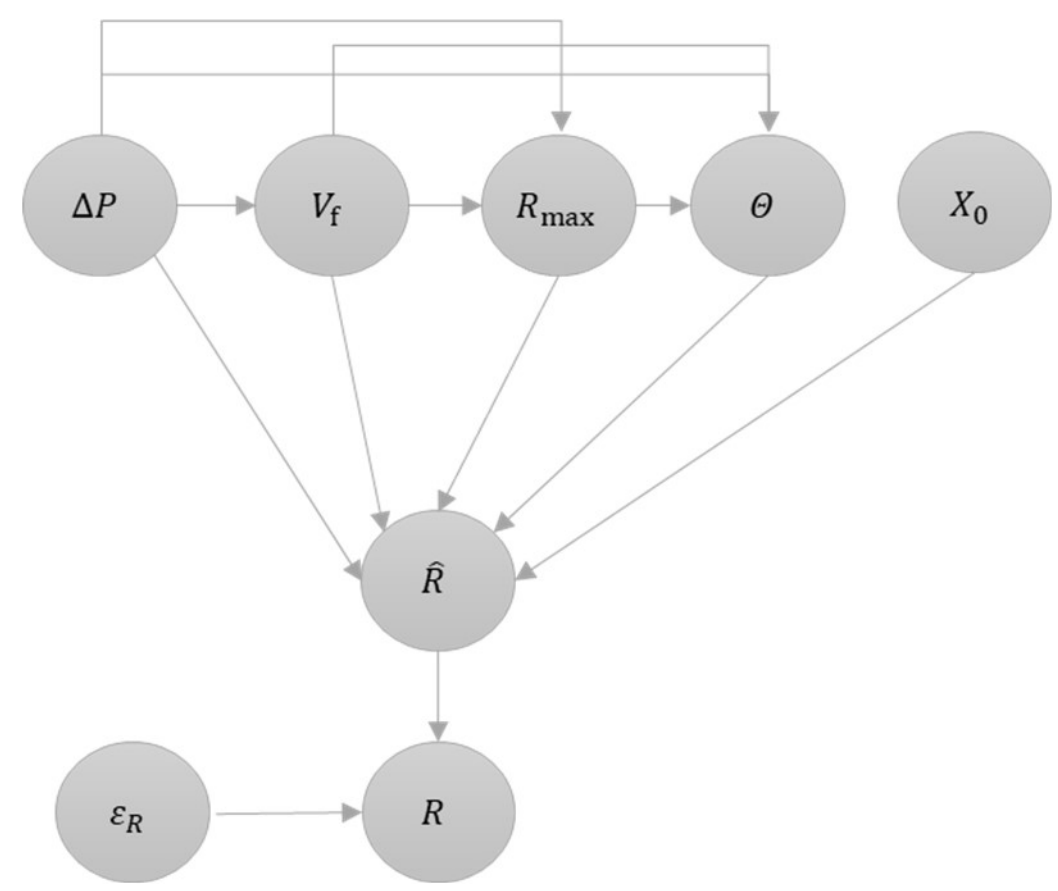

**Fig. 2 BN of TC-induced coastal hazard reflecting the JPM integral**

## 3.3. Structure of BN reflecting multi-hazard JPM analysis

The JPM integral in Equation (2) can be extended to model multiple hazards:

$$\lambda_{R_1>r_1^* \cap R_2>r_2^*} = \lambda \int P(R_1 > r_1^* \ \cap \ R_2 > r_2^* | \mathbf{x}, \varepsilon) \, f_{\mathbf{X}}(\mathbf{x}) f_{\varepsilon}(\varepsilon) d\mathbf{x} d\varepsilon \tag{6}$$

Where $P(R_1 > r_1^* \ \cap \ R_2 > r_2^* | \mathbf{x}, \varepsilon)$ represents the probability that storm response quantities jointly exceed specified thresholds $r_1^*$ and $r_2^*$ (i.e., $R_1 > r_1^*$ and $R_2 > r_2^*$), given TC parameter forcing vector **x** and error $\varepsilon$. All other aspects are defined consistently with the explanation in Section 3.2. The associated BN is shown in Fig. 3, a joint node $J_{R_1 \cap R_2}$ is added to compute the joint probability of different combinations of discretized bins of node $R_1$ and node $R_2$. The node $J_{R_1 \cap R_2}$ is assigned a CPT of $p\left(j_{R_1 \cap R_2} | \, r_1, r_2\right)$, which provides the probability that the joint hazard falls into each of the $k = 1, \ldots, N_{R_1} * N_{R_2}$ mutually exclusive combinations of the states of the discretized bins of $R_1$ and $R_2$. That is, if the $l$th state of $J_{R_1 \cap R_2}$ represents the combination of $r_{1,i}$ and $r_{2,j}$, then:

$$p(J_{R_1 \cap R_2} = j_{R_1 \cap R_2, k} | r_1, r_2) = \begin{cases} 1, if \ \left[r_{1,i}^- \leq R_1 < r_{1,i}^+ \ \cap \ r_{2,i}^- \leq R_2 < r_{2,i}^+\right] \\ 0, Otherwise \end{cases} \tag{7}$$

where $r^-$ and $r^+$ are the lower and upper boundaries of the corresponding bin interval.

The BN in Fig. 3 has a joint distribution specified as:

$$p(\boldsymbol{\omega}) = p\left(j_{R_1 \cap R_2} \mid r_1, r_2\right) \times p\left(r_1 \middle| \hat{r}_1, \varepsilon_{R_1}\right) \times p\left(\varepsilon_{R_1}\right) \times p\left(r_2 \middle| \hat{r}_2, \varepsilon_{R_2}\right) \times p\left(\varepsilon_{R_2}\right) \\ \times p(\hat{r}_1 | \delta p, v_\mathrm{f}, r_\mathrm{max}, \theta, x_0) \times p(\hat{r}_2 | \delta p, v_\mathrm{f}, r_\mathrm{max}, \theta, x_0) \times p(\theta | \delta p, v_\mathrm{f}, r_\mathrm{max}) \\ \times p(r_\mathrm{max} | \delta p, v_\mathrm{f}) \times p(v_\mathrm{f} | \delta p) \times p(\delta p) \times p(x_0) \tag{8}$$

Similar to Equation (5), the JPM integral portion of multi-hazard in Equation (6) can be approximately computed as:

$$\int P(R_1 > r_1^* \,\cap\, R_2 > r_2^* | \mathbf{x}, \varepsilon)\, f_\mathbf{X}(\mathbf{x}) f_\varepsilon(\varepsilon) d\mathbf{x} d\varepsilon \approx \sum_{j_{R_1 \cap R_2, k} \text{ s.t.} r_1 > r_1^* \cap r_2 > r_2^*} p(j_{R_1 \cap R_2, k}) \tag{9}$$

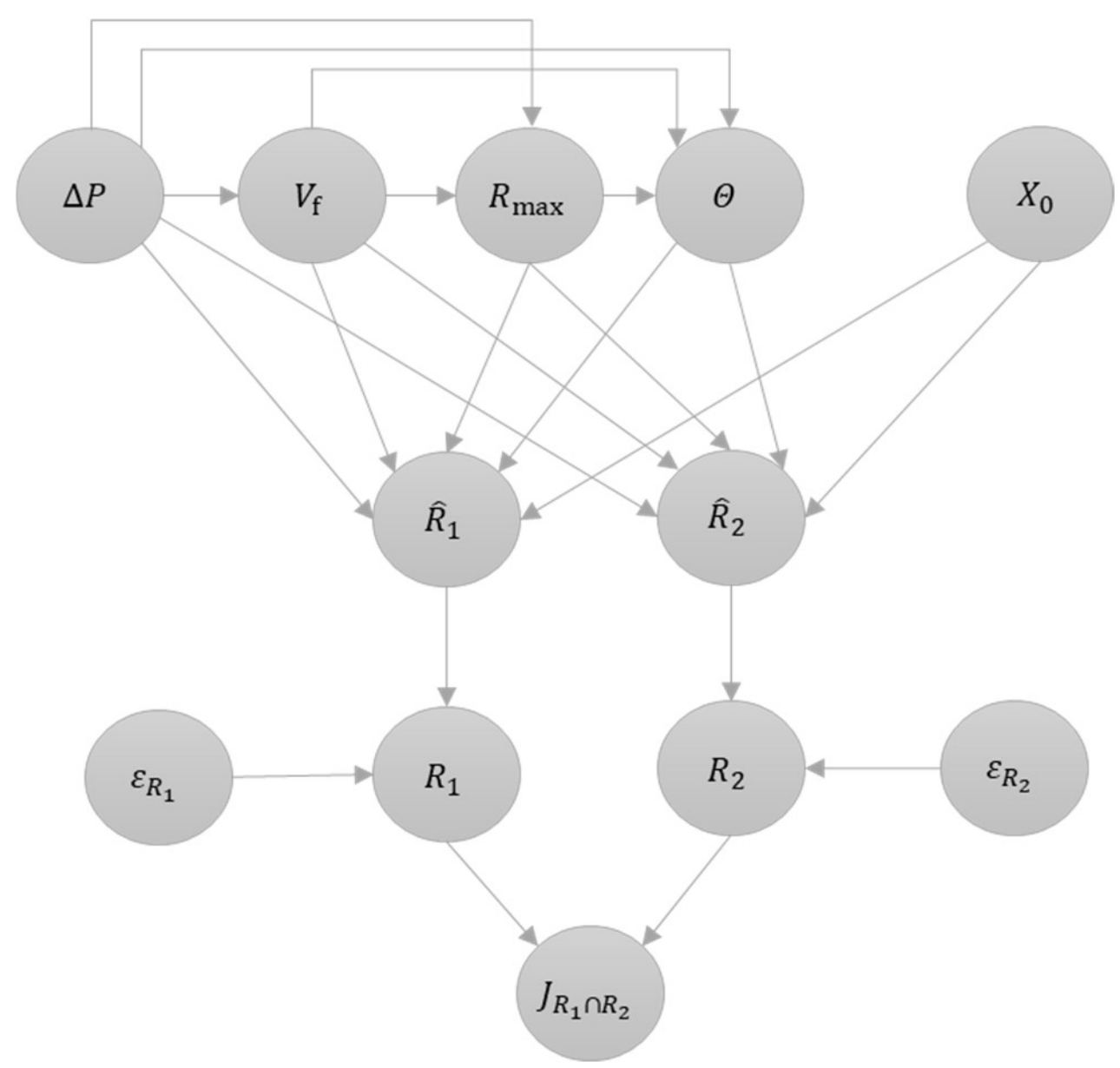

**Fig. 3 BN of multiple TC-induced coastal hazards**

## 4. Implementation of BN for multi-hazard JPM analysis

In the previous section, we described how the JPM leads to the structure of the BN. Once the structure of the BN is defined, the CPTs associated with each node must be specified. While the overall structure of the BN will be generally consistent for any application of the BN to multi-hazard assessment, the CPTs assigned to the nodes will be specific to a study location.

To demonstrate the implementation of the BN, a case study was performed in the Greater New Orleans region in Louisiana (USA). The study center was set at latitude = 29.58 and longitude = -89.54 with a 600 km radius capture zone. This location is shown in Fig. 4 (a) and will be referred to as the coastal reference location (CRL) throughout this paper. Within the study region, two locations are selected to calculate TC-induced coastal hazards: one site within the Mississippi River (Site MR) and one site along the Gulf coast (Site GC). The two study sites are identified in Fig. 4 (b). The coastal hazards represented by $R_1$ and $R_2$ are peak storm surge and TC-induced storm total rainfall (STR).

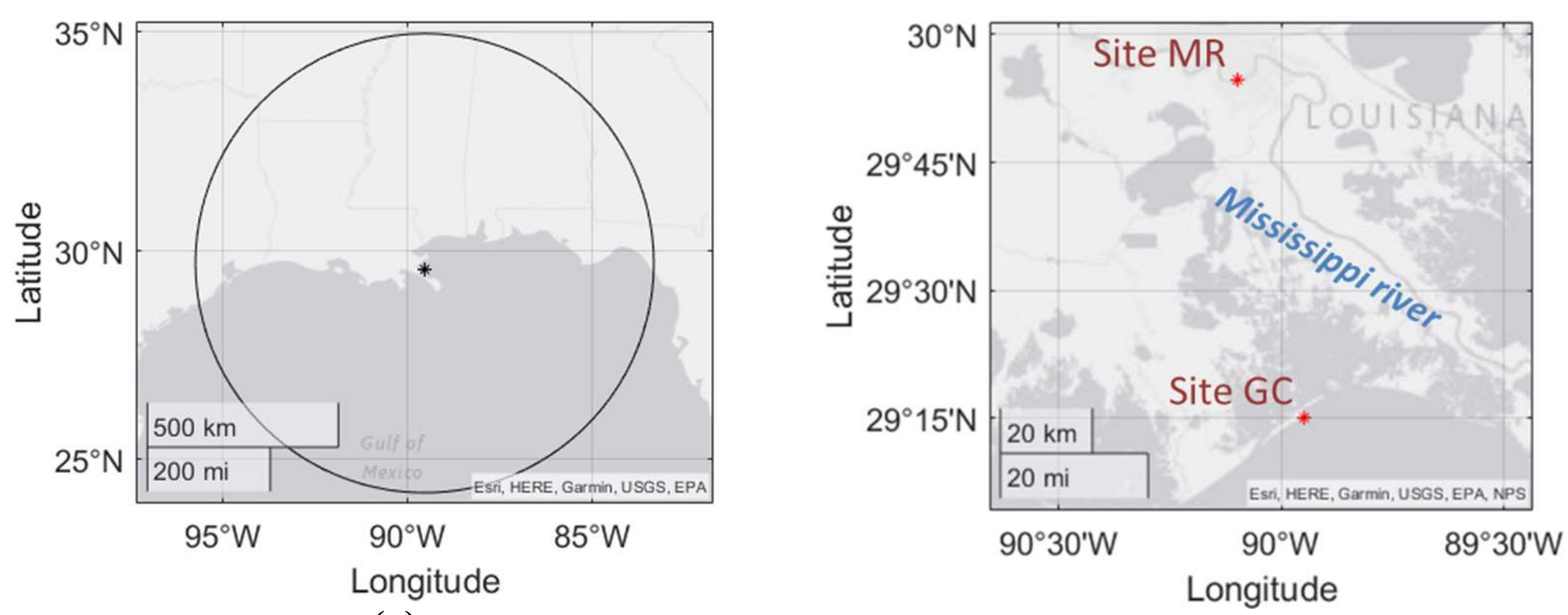

(a) (b)

**Fig. 4 (a) CRL [29.58, -89.54] and capture zone; (b) Study sites, Site MR [29.91, -90.10]; Site GC [29.20, -90.00]**

An independent function library, SMILE, and its graphical user interface software GeNIe (BayesFusion 2023) is used in this study for BN construction and computation. The BN configuration is adapted further to leverage the U.S. Army Corps of Engineers' (USACE) Coastal Hazard System (CHS; https://chs.erdc.dren.mil) Probabilistic Framework (PF) with Joint Probability Method Augmented by Metamodel Prediction (JPM-AMP) (Nadal-Caraballo et al. 2020, 2022a; b). A node $I$ is added to the BN to reflect recent modeling in which the probability distribution of TC parameters is defined separately for low-, medium-, and high-intensity (LI, MI, and HI) TCs, with partitioning based on $\Delta P$.[1] The associated joint distribution over all random variables is then defined as:

$$p(\boldsymbol{\omega}) = p\left(j_{R_1 \cap R_2} \,|\, r_1, r_2\right) \times p\left(r_1 \middle| \widehat{r_1}, \varepsilon_{R_1}\right) \times p\left(\varepsilon_{R_1}\right) \times p\left(r_2 \middle| \widehat{r_2}, \varepsilon_{R_2}\right) \times p\left(\varepsilon_{R_2}\right) \quad (10)$$
$$\times\, p(\hat{r}_1 | \delta p, v_{\mathrm{f}}, r_{\max}, \theta, x_0) \times p(\widehat{r_2} | \delta p, v_{\mathrm{f}}, r_{\max}, \theta, x_0) \times p(\theta | \delta p, v_{\mathrm{f}}, r_{\max})$$
$$\times\, p(r_{\max} | \delta p, v_{\mathrm{f}}) \times p(v_{\mathrm{f}} | \delta p) \times p(\delta p | i) \times p(x_0) \times p(i)$$

Additionally, a joint node $J_{\mathrm{TC}}$ is added and assigned as a "dummy node"[2] that is a child of TC parameter nodes for the purpose of computing the joint probability of TCPCs using a clustering algorithm provided by SMILE (Dawid 1992; Jensen 1990; Lauritzen and Spiegelhalter 1988; BayesFusion 2023). In this case study, node $J_{R_1 \cap R_2}$ is designated with its states representing different combinations of discretized bins of node $R_1$ and node $R_2$, and an identity matrix is assigned as its CPT. The node $J_{R_1 \cap R_2}$ is able to return the joint probability of different combinations of discretized bins of node $R_1$ and node $R_2$ and can be used to generate multi-hazard JPM analysis as presented in Section 5.1. The node $E_{R_1}$ and $E_{R_2}$ are designated for entering evidence (i.e., setting nodes to a particular state) to generate results (deaggregation results) under evidence, as presented in Section 5.2. Each of these nodes contains two states: "true" and "false." For the node $E_{R_i}$, the state is true when $R_i > r_i$, where $r_i$ is a specified threshold of interest for $R_i$. The configuration of BN for this case study implementation is showed in Fig. 5.

[1] LI TCs (8 hPa $\leq \Delta P <$28 hPa), MI TCs (28 hPa $\leq \Delta P <$48 hPa), and HI TCs (48 hPa $\leq \Delta P$).

[2] The term "dummy node" is used when a node is added to the BN for convenience or to enable quantification of a specific quantity but does not have a physical meaning within the BN.

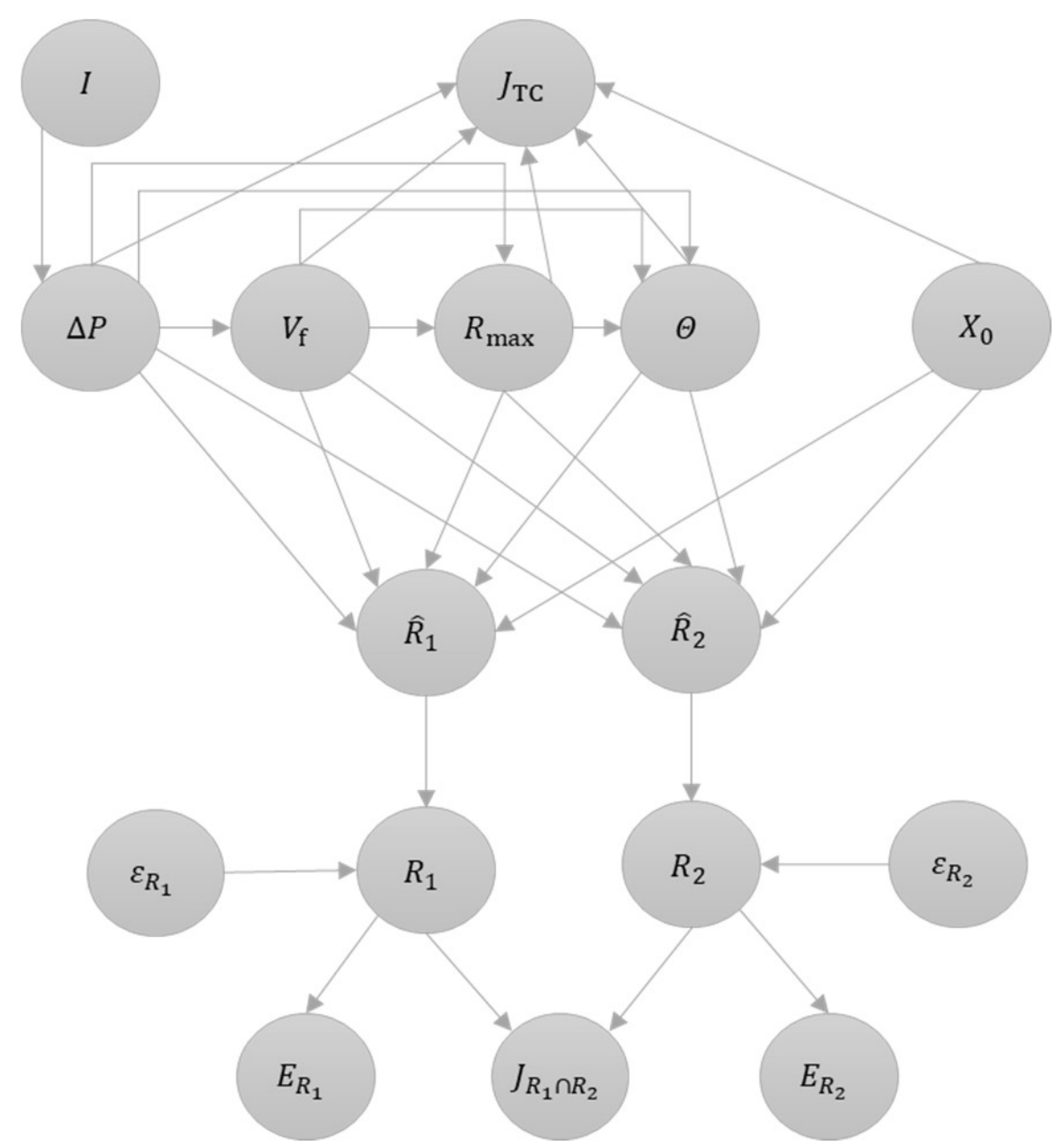

**Fig. 5 BN configuration of case study**

The conditional probability distributions shown in Equation (10) must be defined, discretized, and converted into CPTs to enable quantification within the BN. We begin with an explanation of the process for defining the conditional distributions $p(\theta|\delta p, v_{\mathrm{f}}, r_{\max})$, $p(r_{\max}|\delta p, v_{\mathrm{f}})$, $p(v_{\mathrm{f}}|\delta p)$, $p(\delta p|i)$, $p(i)$, and $p(x_0)$.

First, we define the continuous joint distribution, $f(\delta p, r_{\max}, v_{\mathrm{f}}, \theta)$, associated with the TC parameter nodes $\Delta P, R_{\max}, V_{\mathrm{f}}$, and $\Theta$. The continuous joint distribution is then discretized and used to define the BNs' CPTs. The assumptions of this case study for constructing the joint distribution $f(\delta p, r_{\max}, v_{\mathrm{f}}, \theta)$ are based on the approach of Nadal-Caraballo et al. (2020). The approach integrates historical storm data from the HURDAT2 (Landsea and Franklin 2013; NOAA 2022) and EBTRK (Demuth et al. 2006; RAMMB/CIRA 2021) databases. Data imputation methods provide a strategy to address missing observations in historical datasets (Agrawal et al. 2025; Liu et al.

2024c). In this work, missing values for central pressure and $R_{\max}$ are imputed using Gaussian process models (Liu et al. 2024c). Using this dataset, along with the CRL and the defined capture zone, statistical samples are generated by extracting the maximum-intensity track point for each storm passing through the capture zone. A distance-weighted adjustment is applied following the method introduced in USACE CHS-PF (Nadal-Caraballo et al. 2015). The statistical analysis involves specifying the continuous marginal distributions of TC parameter nodes, followed by the specification of a copula function to capture the dependence structure between variables. Consistent with Nadal-Caraballo et al. (2020), marginal distribution models are constructed for each TC intensity category $i$: $f(\delta p, v_{\mathrm{f}}, r_{\max}, \theta | i)$. The truncated Weibull distribution is used for $\Delta P$ and the lognormal distribution is used for $V_{\mathrm{f}}$ and $R_{\max}$. For the heading direction $\Theta$, a directional storm recurrence rate (DSRR) (Chouinard and Liu 1997) is used to represent its probability model. Leveraging the work of Liu et al. (2024a), the directional weight within the DSSR is a von Mises Kernel function (VKF) (Taylor 2008). A Meta-Gaussian copula model (MGC) is used as the copula function, as proposed by Nadal-Caraballo et al. (2020).

After the continuous joint distribution is specified, it must be discretized before it can be used to define the BNs' CPTs. To explain this process, consider a continuous joint probability density function for a set of $m$ random variables $f(x_1, x_2, \dots x_m)$. Suppose that the domains of each random variable have been discretized into a set of contiguous, mutually exclusive, and collectively exhaustive states. The joint probability defined for a discretized variable combination $\left\{x_1^{[i_1]}, x_2^{[i_2]}, \dots x_m^{[i_m]}\right\}$ can be calculated as:

$$p\left(x_1^{[i_1]}, x_2^{[i_2]}, \dots x_m^{[i_m]}\right) = P\left(x_1^{[i_1-]} < X_1 < x_1^{[i_1+]} \cap \dots \cap x_m^{[i_m-]} < X_m < x_m^{[i_m+]}\right) \quad (11)$$

$$= \int_{x_1^{[i_1-]}}^{x_1^{[i_1+]}} \int_{x_2^{[i_2-]}}^{x_2^{[i_2+]}} \dots \int_{x_m^{[i_m-]}}^{x_m^{[i_m+]}} f(x_1, x_2, \dots x_m) dx_1 dx_2 \dots dx_m$$

where $x_j^{[i_j-]}$ is the lower limit of the discretized bin corresponding to the state $x_j^{[i_j]}$ for $X_j$ and $x_j^{[i_j+]}$ is the corresponding upper limit.

Next, the conditional probability of any subset of discretized states of the variables can be derived from the joint probability as:

$$p\left(x_1^{[i_1]}, x_2^{[i_2]}, \dots x_k^{[i_k]} | x_{k+1}^{[i_{k+1}]}, \dots x_m^{[i_m]}\right) = \frac{p\left(x_1^{[i_1]}, x_2^{[i_2]}, \dots x_m^{[i_m]}\right)}{p\left(x_{k+1}^{[i_{k+1}]}, \dots x_m^{[i_m]}\right)} \quad (12)$$

The above general approach is used in this study to discretize the joint distribution $f(\delta p, v_{\mathrm{f}}, r_{\max}, \theta | i)$ to obtain the conditional discretized distributions $p(\theta | \delta p, v_{\mathrm{f}}, r_{\max})$, $p(r_{\max} | \delta p, v_{\mathrm{f}})$, $p(v_{\mathrm{f}} | \delta p)$, and $p(\delta p | i)$. It is noted that the intensity partitioning associated with the node $I$ is based only on $\Delta P$. For this reason, the CPTs for $\Theta$, $R_{\max}$, and $V_{\mathrm{f}}$ are not expressed as dependent on $I$. The marginal distribution, $p(i)$, of the root node $I$ is based on the relative likelihood of LI, MI, and HI TCs.

The TC landfall location parameter $X_0$ is defined using a horizontal reference line that is set off the coastline and discretized into segments. In the case study, this is represented by a line at latitude=29.5°, extending from longitude=-93° to longitude=-86°, and discretized to yield 14 representatives of $X_0$ with equal-length spacing, as shown in Fig. 6. This approach is implanted here to enable more meaningful outputs of deaggregation related to TC tracks.

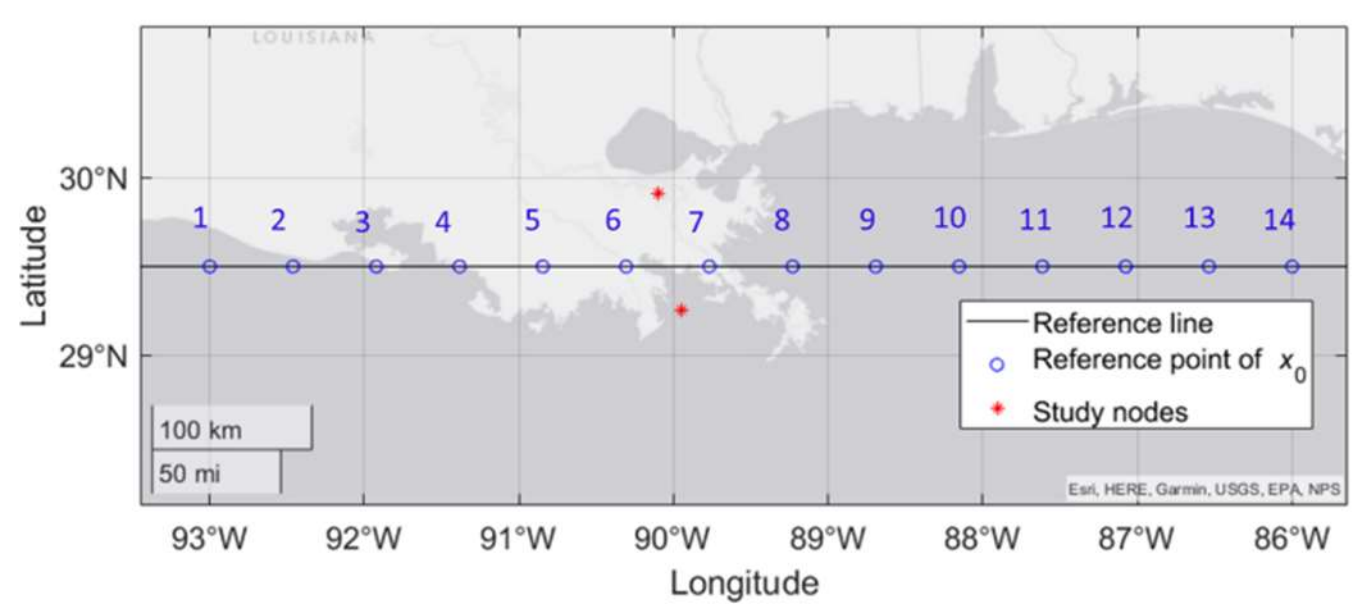

**Fig. 6 Representatives of $X_0$**

Straight line TC tracks are used to determine projected landfall locations for each combination of $X_0$ and discretized $\Theta$ value. The straight line TC tracks and projected landfall locations are plotted in Appendix A.

Next, we describe the process for calculating the CPTs for the nodes $\hat{R}_1$ and $\hat{R}_2$ as a function of its parent nodes; i.e., $p(\hat{r}_1 | \delta p, v_{\mathrm{f}}, r_{\max}, \theta, x_0)$ and $p(\hat{r}_2 | \delta p, v_{\mathrm{f}}, r_{\max}, \theta, x_0)$. The predicted value of the coastal hazard severity measure ($\hat{R}_i$) is a function of TC parameters ($\Delta P$, $V_{\mathrm{f}}$, $R_{\max}$, $\Theta$, $X_0$), as indicated by the arrows from the TC parameters to $\hat{R}_i$. Thus, these CPTs provide the conditional distribution of predicted storm response given a set of storm parameter bins. In general, TC hazard response quantities can be predicted as a function of a TC's characteristics using physical process models, numerical models, empirical models, etc. In recent years, machine-learning-derived surrogate models (e.g., Al Kajbaf and Bensi 2020; Gharehtoragh and Johnson 2024; Jia and Taflanidis 2013; Liu et al. 2024a) have been widely used to efficiently predict TC-induced coastal hazard response quantities (e.g., peak storm surge and STR) as a function of TC parameters $\Delta P$, $V_{\mathrm{f}}$, $R_{\max}$, $\Theta$ and $X_0$. In this study, a Gaussian process regression-based metamodel developed for the Coastal Hazards System–Louisiana (CHS-LA) study (Nadal-Caraballo et al. 2022b) is used as the peak storm surge surrogate model. To model STR, an artificial neural network model[2] is developed using training data derived from the time-series simulations of TC-induced rainfall under the 645 synthetic storms suite developed by the CHS-LA study (Nadal-Caraballo et al. 2022b) using a physics-based TC rainfall model (Lu et al. 2018; Xi et al. 2020; Zhu et al. 2013).

Monte Carlo simulation (MCS) is implemented for the computation of the CPT of nodes $\hat{R}_1$ and $\hat{R}_2$ to reduce the error induced by the discretization of TC parameters (Mohammadi et al. 2023). That is, for each combination of TC parameter bins associated with the parent nodes, $N_{\mathrm{sim}}$ random values are simulated within the bin bounds for each TC parameter following the corresponding marginal distribution truncated to the bin interval. The surrogate model is used to estimate the values of the response quantity for each simulation $i = 1, \dots, N_{\mathrm{sim}}$ and the frequency of simulations falling into each mutually exclusive range associated with the response quantity bin is used to define the CPT.

Then, the actual value of the coastal hazard $R_i$ is computed as the superposition of the predicted value of hazard response and an error term:

$$R_i = \hat{R}_i + \varepsilon_{R_i}\sigma_{c_i}(\hat{R}_i) \tag{13}$$

where $\varepsilon_{R_i}$ is the normalized residuals of $R_i$, which follows a standard Gaussian distribution and $\sigma_{c_i}$ is the standard deviation of a combined uncertainty term, which is severity dependent (i.e., a function of $\hat{R}_i$). The process of computing combined uncertainty is described in the Section 6.3 of Nadal-Caraballo et al. (2022b). MCS is implemented for the computation of the CPT of $R_i$ for the purpose of reducing the error induced by the discretization of TC parameters (Mohammadi et al. 2023).

## 5. Illustration of inference using BN for multi-hazard JPM analysis

[2] The "feedforward net" function from the MATLAB Deep Learning Toolbox (MathWorks 2022) is used to train a multilayer feedforward back-propagation ANN model for predicting STR at study locations. The hyperbolic tangent function is used as the transfer function for hidden layers and the output layer. Bayesian regularization back-propagation is used as the regularization strategy. A leave-one-out cross-validation (LOOCV) is applied to assess the performance of this STR surrogate model, with testing RMSE values of approximately 7-8mm for the study locations.

This section demonstrates how the BN can be used for forward and backward inference. Inference refers to the process of computing the posterior probabilities of nodes within the BN given evidence and propagating information through the BN.

In Section 5.1, the forward propagation capabilities of the BN is used to calculate the marginal and joint distributions of coastal hazard nodes under the no-evidence case (i.e., considering only the initial assumptions used to build the model before entering any evidence in the BN). In Section 5.2, forward inference (i.e., propagation in the direction of the links) and backward inference (i.e., propagation opposite the direction of the links) are used to calculate the distributions of nodes in the BN conditioned on evidence indicating the occurrence of low-exceedance-probability (LEP) coastal hazards, which enables deaggregation analyses. It is noted that all results presented in this section are generated by the BN. The storm recurrence rate, which is necessary for computing the annual frequency of exceedance of hazards and developing hazard curves, is not included in these results. Note in the following result sections, the notation of $H$ (Greek letter eta, with lower case printed as $\eta$) and $P$ are used to represent the variable of storm surge and STR (corresponding to $R_1$ and $R_2$) to help with reading results.

## 5.1. Baseline (Prior) results without evidence

The distributions of coastal response quantities for the no-evidence case is represented by the probability of discretized bins and plotted in Fig. 7 for the two study sites. The stem plot represents the probability (represented on the y-axis as $p(\square)$) of each discretized bin, and dashed curves are created by interpolating the probability of discretized bins to approximate the distribution shapes.

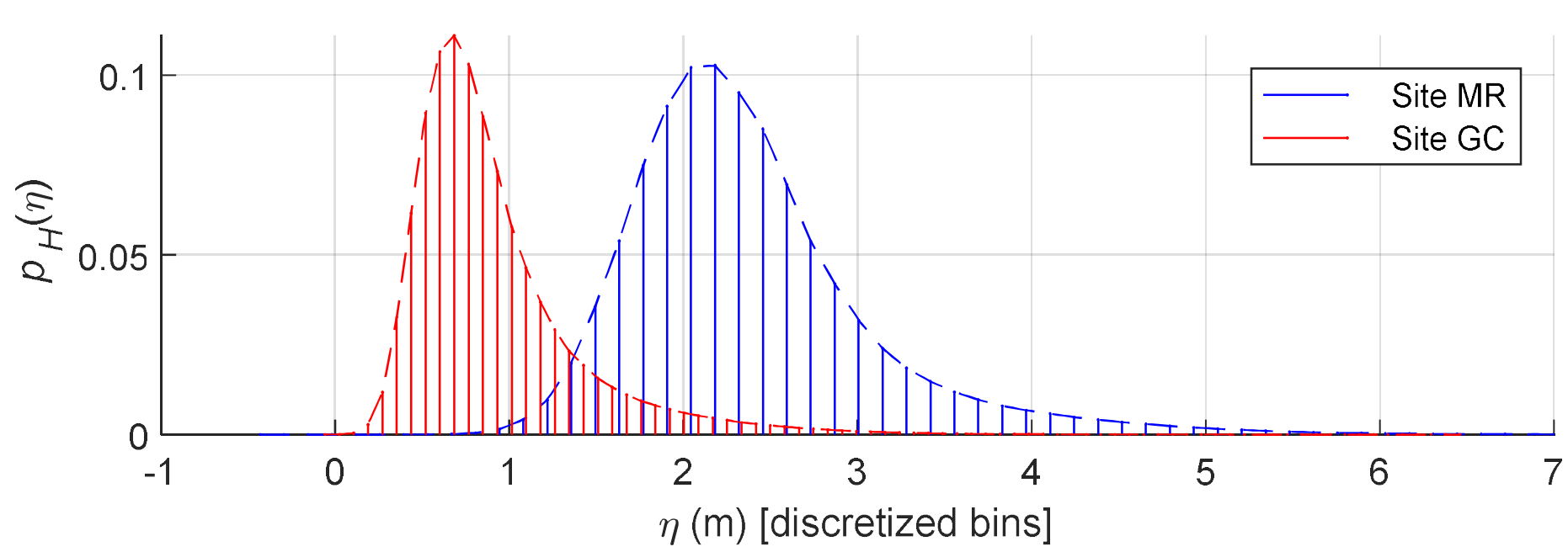

**(a)**

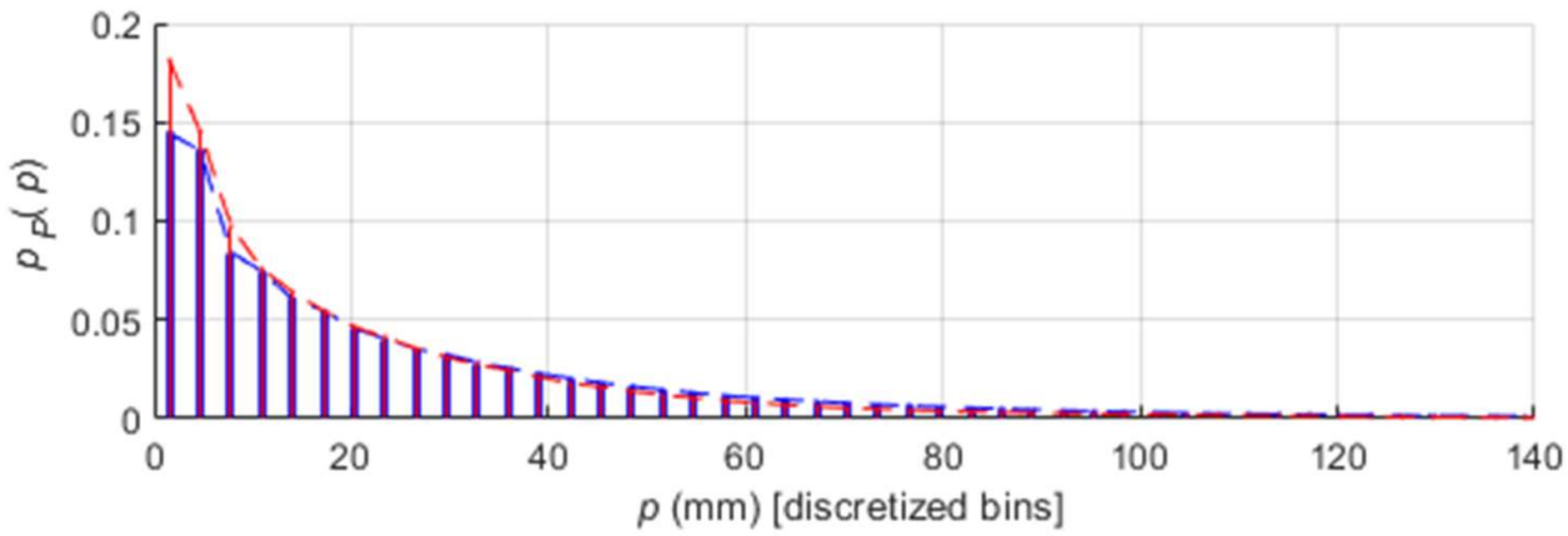

**(b)**

**Fig. 7 Coastal hazard distributions generated from BN; (a) Storm surge; (b) STR**

The exceedance probability is a critical metric for TC-induced coastal hazard JPM integral (Nadal-Caraballo et al. 2020). Leveraging the BN described in Section 4, the joint probabilities of storm surge and STR are computed by node $J_{R_1 \cap R_2}$. Then, the PDF surfaces and joint exceedance probability surfaces are numerically approximated by scaling the computed joint probabilities to ensure that the integrated volume under the PDF surfaces equals one. Fig. 8 plots the approximated joint PDF (represented on the color bar as $f_{H,P}(\eta, p)$) and joint exceedance probability (represented on the color bar as $P(H > \eta \cap P > p)$) surfaces.

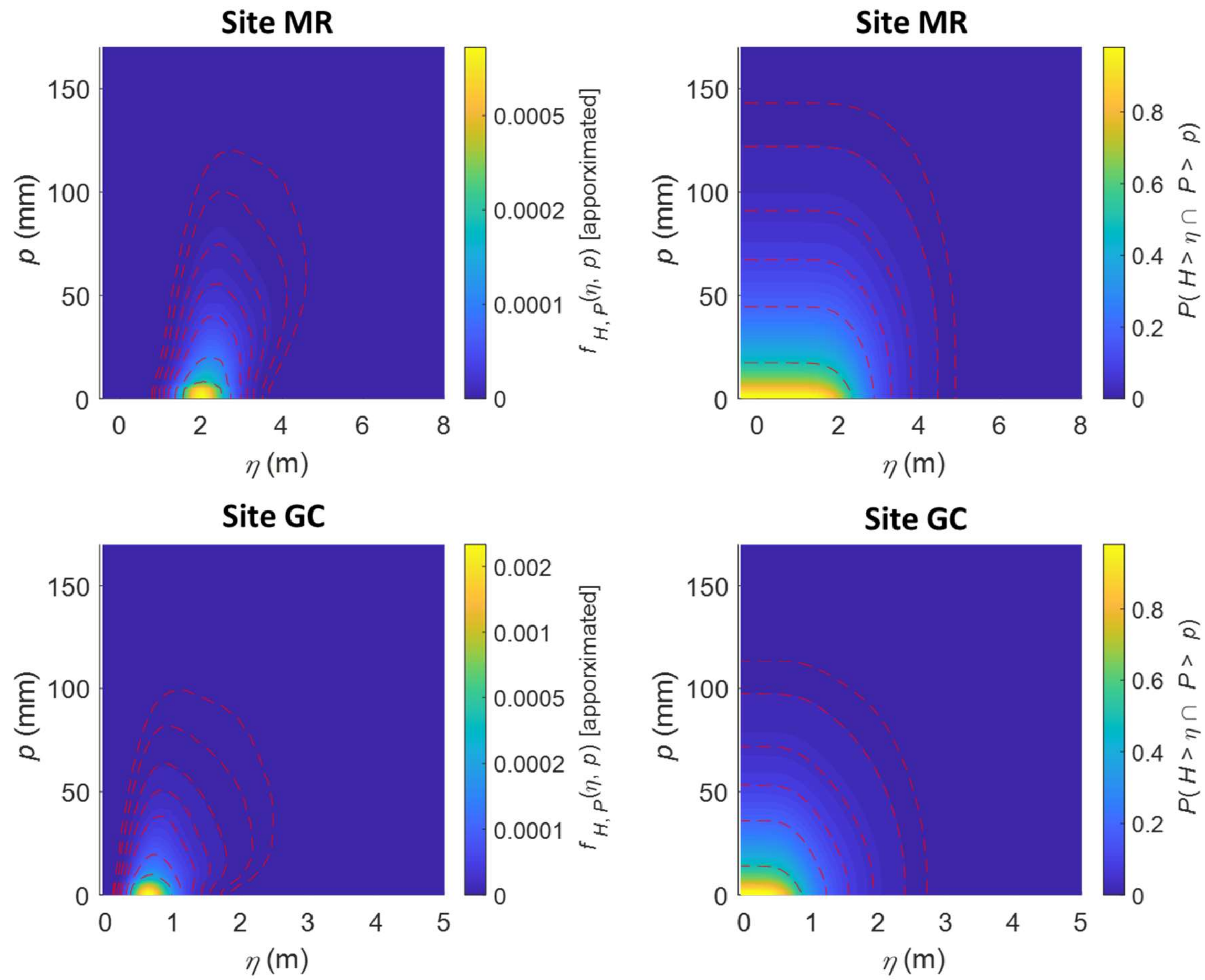

**Fig. 8 Approximated joint PDF and joint exceedance probability of storm surge and STR. Contours are added to help visualize the correlation pattern. Note that color scales are set differently for visualization**

As expected, the peaks of the joint PDF and joint exceedance probability are concentrated at the lower left portion of the surface plot, reflecting the higher likelihood of smaller combinations of response quantities. Moderate correlations between storm surge and STR values are observed from the approximated joint PDF plots. The joint probabilities are employed to approximately compute the correlation coefficients $\rho_{\eta,p}$, which is evaluated as $\rho_{\eta,p}$ = 0.5 at Site MR and $\rho_{\eta,p}$ = 0.48 at Site GC.

## 5.2. Updated (posterior) results with evidence

The forward and backward inference capabilities of BNs are employed to generate the updated (posterior) probability distributions of TC parameters and coastal hazards for each study site under three evidence cases that correspond to events involving the occurrence of LEP coastal hazards. The three evidence cases (noted as EC1~EC3) are:

- EC1: Evidence that storm surge is greater than a specified threshold. The threshold is set as 4.1 m for Site MR and 2.1 m for Site GC. This evidence is entered by defining the CPT of node $E_{R_1}$ using this threshold and then setting the node $E_{R_1}$ to the "true" state.
- EC2: Evidence that STR is greater than a specified threshold. The threshold is set as 107.8 mm for Site MR and 85.9 mm for Site GC. This evidence is entered by defining the CPT of node $E_{R_2}$ using this threshold and then setting the node $E_{R_2}$ to the "true" state.
- EC3: Evidence that storm surge is greater than a specified threshold and STR is greater than a specified threshold (i.e., this evidence case combines EC1 and EC2). This evidence is entered by setting both the node $E_{R_1}$ and $E_{R_2}$ to the "true" state.

The EC1 and EC2 thresholds are selected based on the value of the severity measure that has a 0.03 exceedance probability under the no-evidence case for each site. These evidence cases can be used to identify TCPCs that are most likely to have contributed to the observed large values of storm surge and/or STR (i.e., dominant TC scenarios for severe coastal hazards).

Fig. 9 shows the probabilities (represented on the y-axis as $p(\square)$) of $\Delta P$, $V_{\mathrm{f}}$ and $R_{\max}$. The dashed curves are created by interpolating the discretized bins' probabilities to approximate the distributions' shapes.

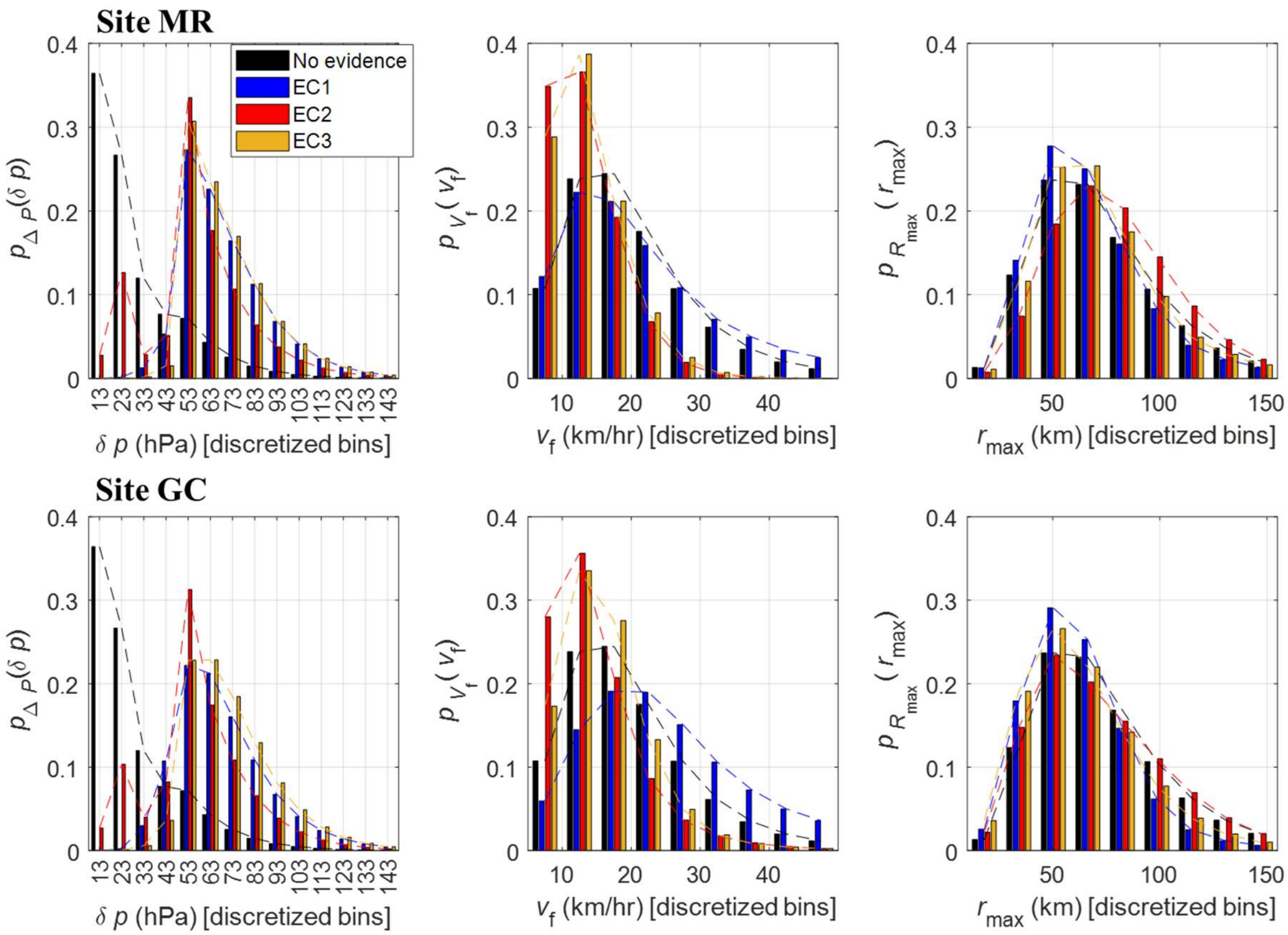

**Fig. 9 Distributions of $\Delta P$, $V_{\mathrm{f}}$ and $R_{\max}$ under EC1 to EC3 at Site MR (row 1) and Site GC (row 2)**

EC1 (high storm surge) shifts the distribution of $\Delta P$ toward higher values, reflecting the higher likelihood of larger $\Delta P$ values given high storm surge events. EC2 (high STR) is associated with a moderately bimodal distribution with a lower likelihood of large $\Delta P$ values than EC1. However, EC3 (high storm surge and STR) results in a more substantial shift toward higher $\Delta P$ than observed under EC1 and EC2 individually.

EC1 (high storm surge) shifts the distribution of $V_{\mathrm{f}}$ rightward, which reflects the higher likelihood of faster-moving storms given high storm surge events. On the other hand, EC2 (high STR) shifts the distribution of $V_{\mathrm{f}}$ leftward, which reflects the higher likelihood of slower-moving storms given high STR events. This can be explained by recognizing that a slower-moving storm gives more time for rainfall accumulation. EC3 (high storm surge and STR) results in a moderated shift between the two evidence cases. The impact of evidence on the distribution of $R_{\max}$ is relatively small when compared with $\Delta P$ and $V_{\mathrm{f}}$.

The effects of evidence on the distribution of $X_0$ and $\Theta$ are considered together because they reflect TC track information. In each case, the $J_{\mathrm{TC}}$ node in the BN, as described in Section 4, is used to compute the joint probabilities of all discretized TCPCs. The joint distribution of TC track parameter combinations (pair-wise joint probability of $X_0$ and $\Theta$ ) for each case is computed by integrating the joint probabilities computed by the $J_{\mathrm{TC}}$ node in the BN over $\Delta P$, $V_{\mathrm{f}}$ and $R_{\max}$. Fig. 10 shows the pair-wise joint probability plot of each case, with joint probabilities indicated on the color bar as $p(\square)$. The no-evidence case is plotted for comparison and shows the equal probability assumption for $X_0$ and the marginal distribution shape for $\Theta$.

From Fig. 10, it can be observed that the input of evidence notably affects the pair-wise joint probability of $X_0$ and $\Theta$ (compared with the no-evidence case). Under EC1, high storm surge values at Site MR lead to a distribution with a mode associated with northwesterly tracks crossing near the midpoint of the $X_0$ reference line, while the distribution is a bit more dispersed when a high storm surge is observed at Site GC. This pattern is also moderately reflected under EC2. However, the TC tracks in EC3 show a more concentrated dominant area than EC1 and EC2, reflecting a smaller range of dominant tracks, given that large values of both storm surge and STR are observed.

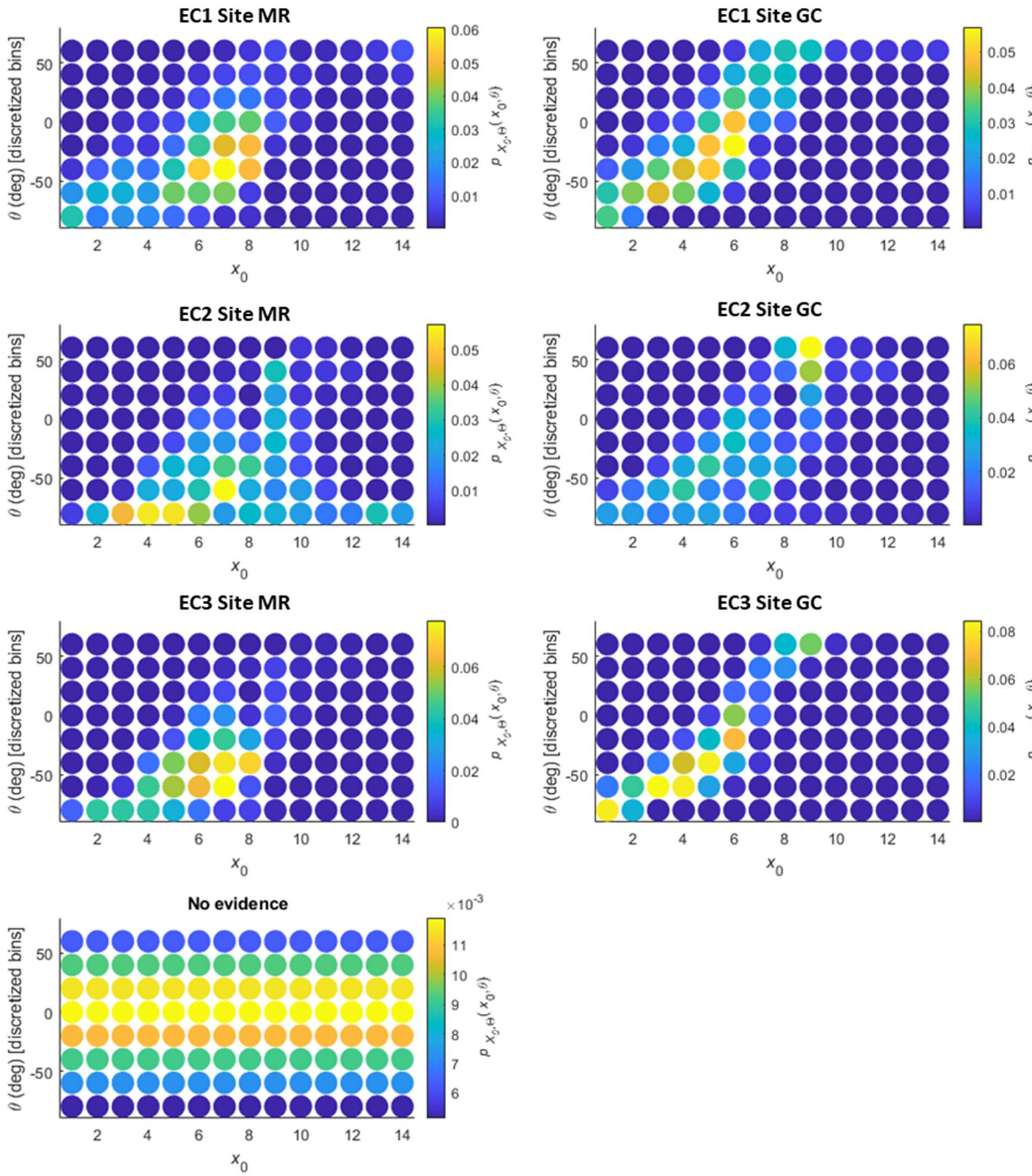

**Fig. 10 Pair-wise joint probability of $X_0$ and $\Theta$ under each evidence case**

The contributions of $\Delta P$, $V_\mathrm{f}$ and $R_\mathrm{max}$ along with the TC track variables are visualized in Fig. 11 to Fig. 13 as 3D stacked bar plots with partitions of $\Delta P$, $V_\mathrm{f}$ and $R_\mathrm{max}$ identified with different colors (joint probabilities are indicated on bar height as $p(\square)$).

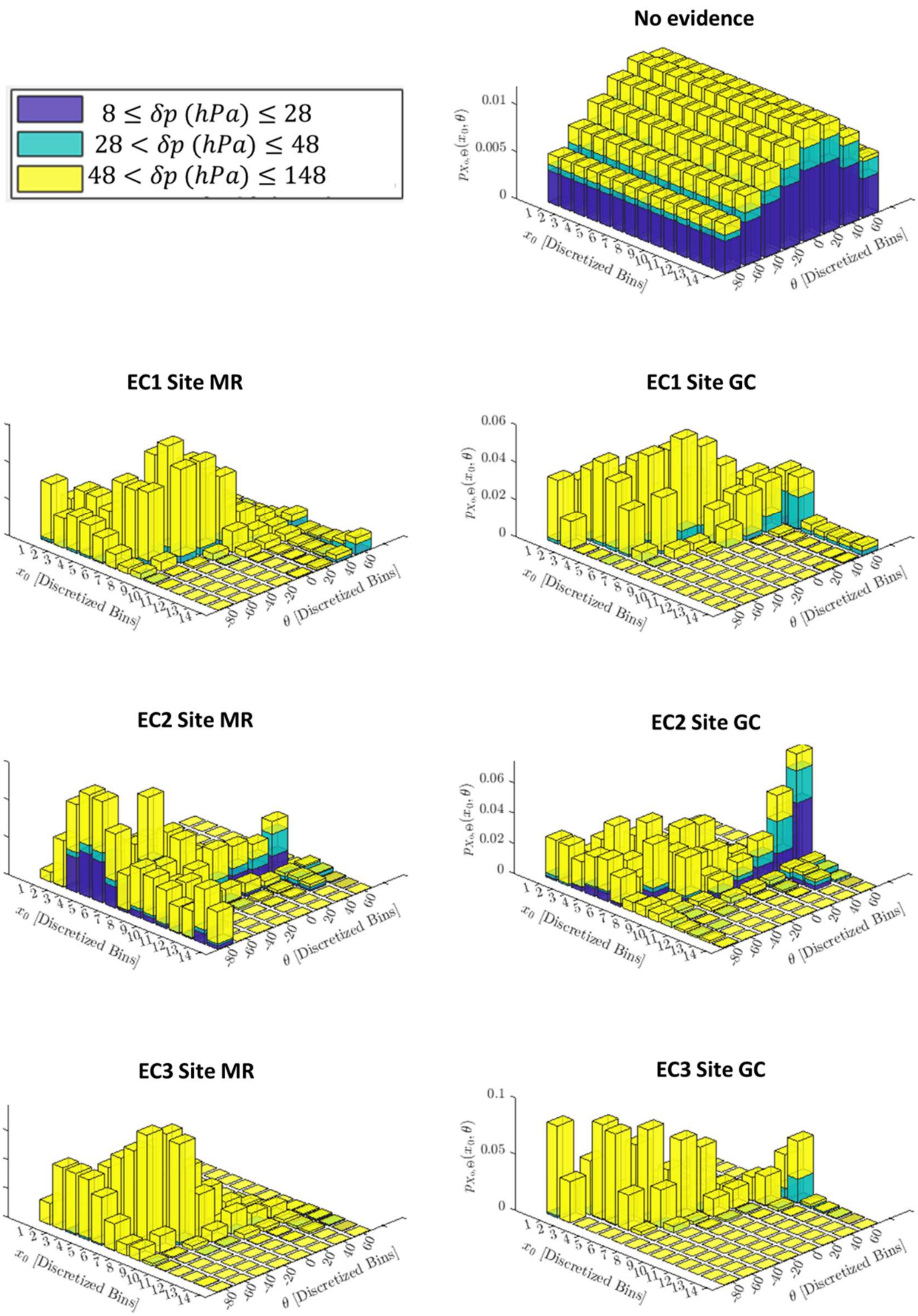

**Fig. 11 3D stacked bar plot of TC track information with $\Delta P$ identified by colors**

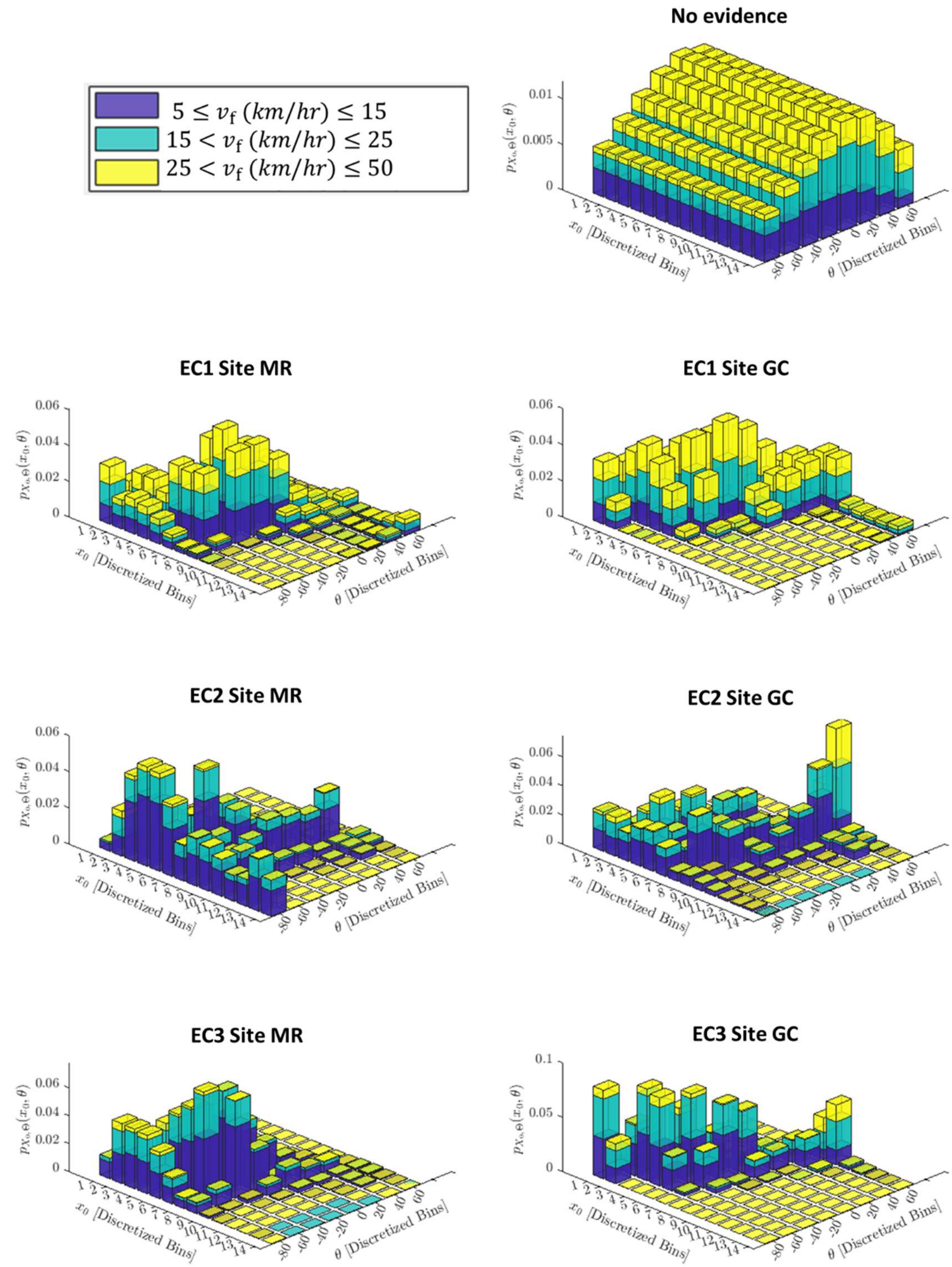

**Fig. 12 3D stacked bar plot of TC track information with $V_f$ identified by colors**

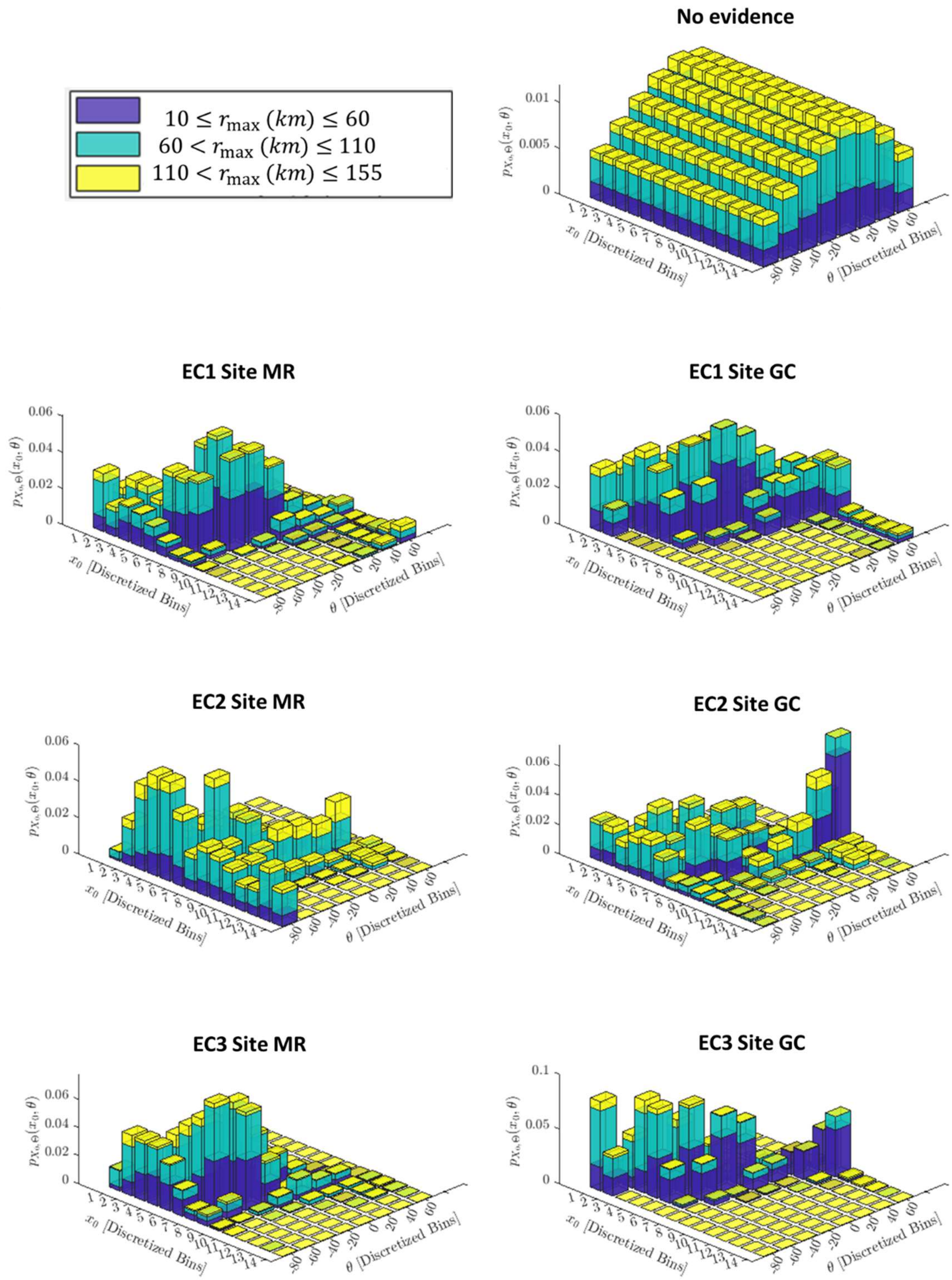

**Fig. 13 3D stacked bar plot of TC track information with $R_{\max}$ identified by colors**

From Fig. 11, it can be observed that TCs with high $\Delta P$ ($\Delta P$ >48 hPa) dominate most evidence cases. This is different with the no-evidence case, which is dominated by lower $\Delta P$ values. Despite the general dominance of high $\Delta P$ values across different evidence cases, EC2 is associated with a higher likelihood of TCs with low $\Delta P$ ($\Delta P$ <48 hPa). Fig. 12 shows that the dominant range of $V_{\mathrm{f}}$ can be notably different under different evidence cases. The likelihood of faster-storms ($V_{\mathrm{f}}$ >25 km/hr) under EC1 is greater than under EC2, which is consistent with the observation from Fig. 9, which implies that a faster-moving storm is likely to generate a higher storm surge while a slower-moving storm can give more time for rainfall accumulation, thereby generating a higher STR. Fig. 13 shows that, for Site MR, the contribution of smaller $R_{\max}$ ($R_{\max}$ <60 km) to dominant TC tracks is greater under EC1 than under EC2, however, this difference is not observed in Site GC for the two most dominant TC tracks with $\Theta \geq 40°$ (the two bars at the right end of EC2 Site GC plot).

Leveraging backward and forward propagation in the BN, information updating is performed to calculate the distribution of one coastal hazard given evidence associated with the occurrence of severe values of another coastal hazard. That is, the distributions of STR are computed under EC1 (high storm surge), and the distributions of storm surge are computed under EC2 (high STR). Fig. 14 shows the resulting distributions, with probability indicated on the y-axis as $p(\square)$. Note that the plotted range of variables in each figure is different.

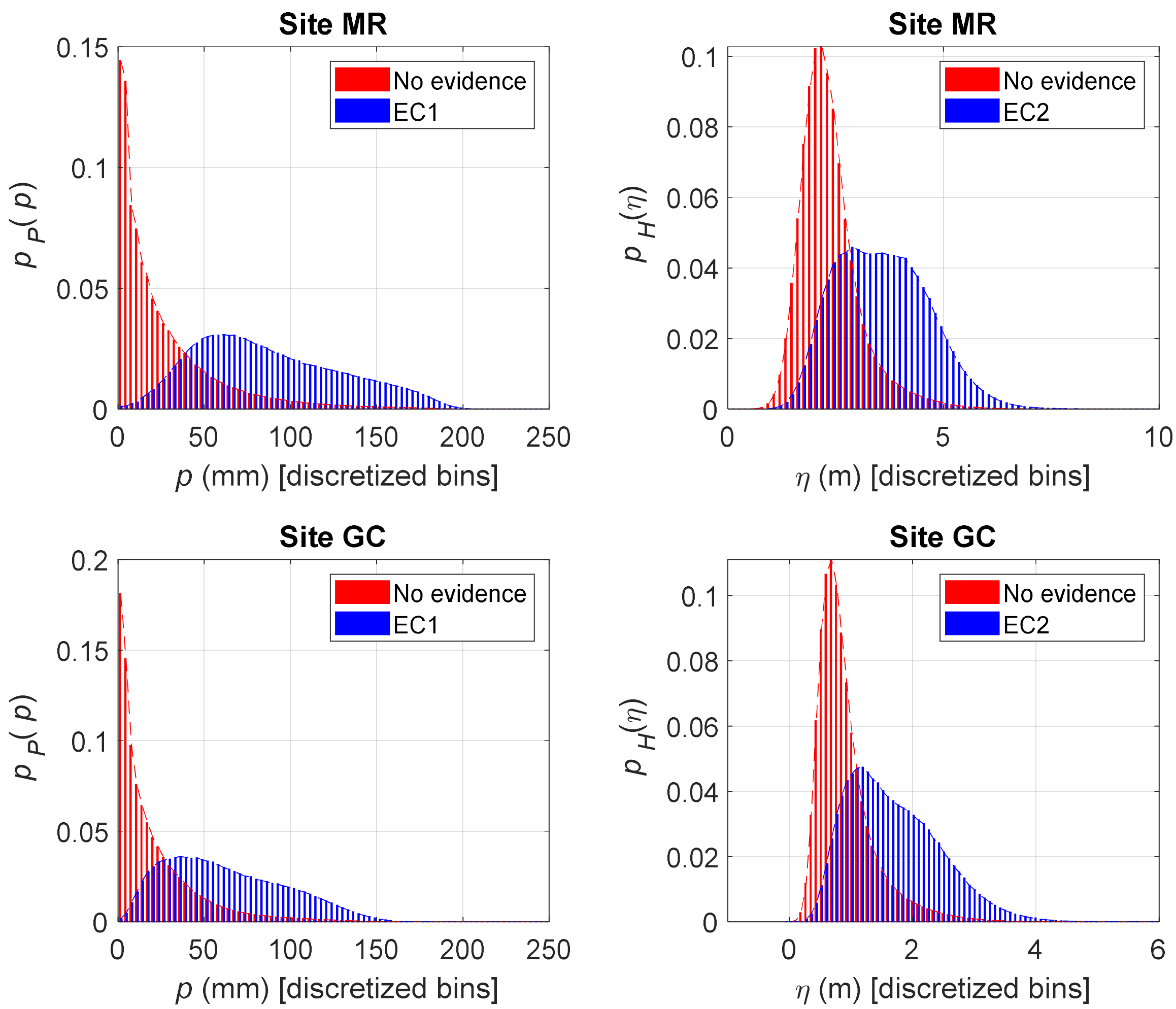

**Fig. 14 Distribution of coastal hazards under no evidence and an evidence case involving the occurrence of an LEP event of another coastal hazard**

It is observed that the peak probabilities of storm surge and STR shift rightward under the evidence case of a LEP event of the other coastal hazard, which can be explained by the connection between intense storms and severe coastal hazards. Namely, EC1 (high storm surge) is associated with a higher likelihood of high STR, while EC2 (high STR) is associated with a higher likelihood of high storm surge at both sites.

## 6. Conclusion

This study implements the JPM for multi-hazard assessment using BNs, offering several advantages not readily achievable with the conventional JPM approach. We present the proposed BN structure and describe the implementation for a case study near New Orleans, LA. Using the case study BN, we demonstrate how the backward inference capabilities of BNs can be used to perform deaggregation and identify dominant sets of storm parameters that lead to severe local hazards, representing a new capability within the PCHA.

Specifically, by leveraging the BN inference capabilities, we demonstrate how to compute the updated (posterior) distributions of TC parameters under evidence cases involving LEP coastal hazards. This can provide insights regarding dominant TCPCs, including track geometries defined by landfall location and heading. Additionally, BN inference is used to examine the impact of observations related to one type of hazard (e.g., storm surge) on the conditional distribution of another (e.g., STR). These results help identify the primary drivers of coastal hazards at a given location and can inform efforts to refine the selection of storm parameter sets.

The updated (posterior) distributions under evidence cases can be used to risk-inform model refinement and augmentation activities by focusing efforts on the events most likely to have occurred given observations of consequential effects. The proposed approach can enhance the ability to leverage JPM results for local or site-scale studies and to inform the extension of single-hazard analyses to account for the impact of compound coastal hazards.

## Acknowledgments

The development of the USACE's CHS (https://chs.erdc.dren.mil), the CHS Probabilistic Framework (CHS-PF), the CHS Compound Coastal-Inland Framework (CHS-CF), the Joint Probability Method Augmented by Metamodel Prediction (JPM-AMP), and related datasets have been funded in part by CHS multi-agency initiative (USACE's Civil Works R&D Programs, FEMA, U.S. Nuclear Regulatory Commission). This study was supported by the USACE's Engineer Research and Development Center, Coastal and Hydraulics Laboratory (ERDC-CHL) under contract award number ID W912HZ20C0050.

## Disclaimer

All opinions expressed in this paper are the authors' and do not necessarily reflect the policies and views of the research sponsor (USACE) or any other organization.

**Competing Interests**

The authors have no relevant financial or non-financial interests to disclose.

**Author Contributions**

ZL: Conceptualization, Methodology, Software, Investigation, Formal analysis, Data Curation, Visualization, Writing—Original Draft. MC: Conceptualization, Methodology, Visualization, Project administration, Writing—Review & Editing, Supervision, Validation. NNC: Data Curation, Writing—Review & Editing, Supervision, Funding Acquisition. MY: Data Curation, Writing—Review & Editing. MB: Conceptualization, Methodology, Software, Investigation, Formal analysis, Visualization, Supervision, Writing—Original Draft, Writing—Review & Editing.

# Appendix A Supplemental Information for Statistical Analysis and Discretization

This appendix provides details of the statistical analysis and discretization of the case study. Table 2 provides the fitted marginal distribution parameters. Fig. 15 to Fig. 17 presents the fitted marginals and correlation information among TC atmospheric parameters. In each figure, the diagonals are histogram plots of statistical sampling data with fitted PDF curves and cumulative density function curves. The marginal distribution analysis is based on those recent implementations of the USACE's CHS-PF with JPM-AMP (Nadal-Caraballo et al. 2020; 2022a; b) as described in Section 4. The off-diagonals are scatter plots of two TC atmospheric parameters. On top of each off-diagonal scatter plot, Pearson's correlation coefficient $\rho$ are reported, which are derived from the measured Kendall's $\tau$ shown in parentheses (Fang et al. 2002). The derived correlation matrix is used as the parameter of MGC. Table 4 provides discretization information for variables (nodes) in the BN as depicted in Fig. 5. Fig. 18 presents the straight line TC tracks and projected landfall locations as described in Section 4.

**Table 2: Marginal distribution parameters**

| Variable | Name of distribution | Truncate | Distribution parameters | |
|---|---|---|---|---|
| $\Delta P$ | Weibull | LI：[8,28] | | $a = 25.79$ |
| | | MI：[28,48] | | $b = 1.197$ |
| | | HI：[48,148] | | |
| $V_{\mathrm{f}}$ | Lognormal | | LI： | $\lambda = 2.848$; $\zeta = 0.4857$ |
| | | | MI： | $\lambda = 2.970$; $\zeta = 0.3518$ |
| | | | HI： | $\lambda = 3.006$; $\zeta = 0.5465$ |
| $R_{\max}$ | Lognormal | | LI： | $\lambda = 4.307$; $\zeta = 0.4170$ |
| | | | MI： | $\lambda = 4.097$; $\zeta = 0.3597$ |
| | | | HI： | $\lambda = 4.009$; $\zeta = 0.4276$ |
| $\Theta$ | Scaled VKF DSRR | [-180,180] | $h_d = 200$ $k$m; $\kappa = 4$ | |

Note: $\Theta$ is measured clockwise within the range of [-180, 180] degrees using north as the zero-degree direction.

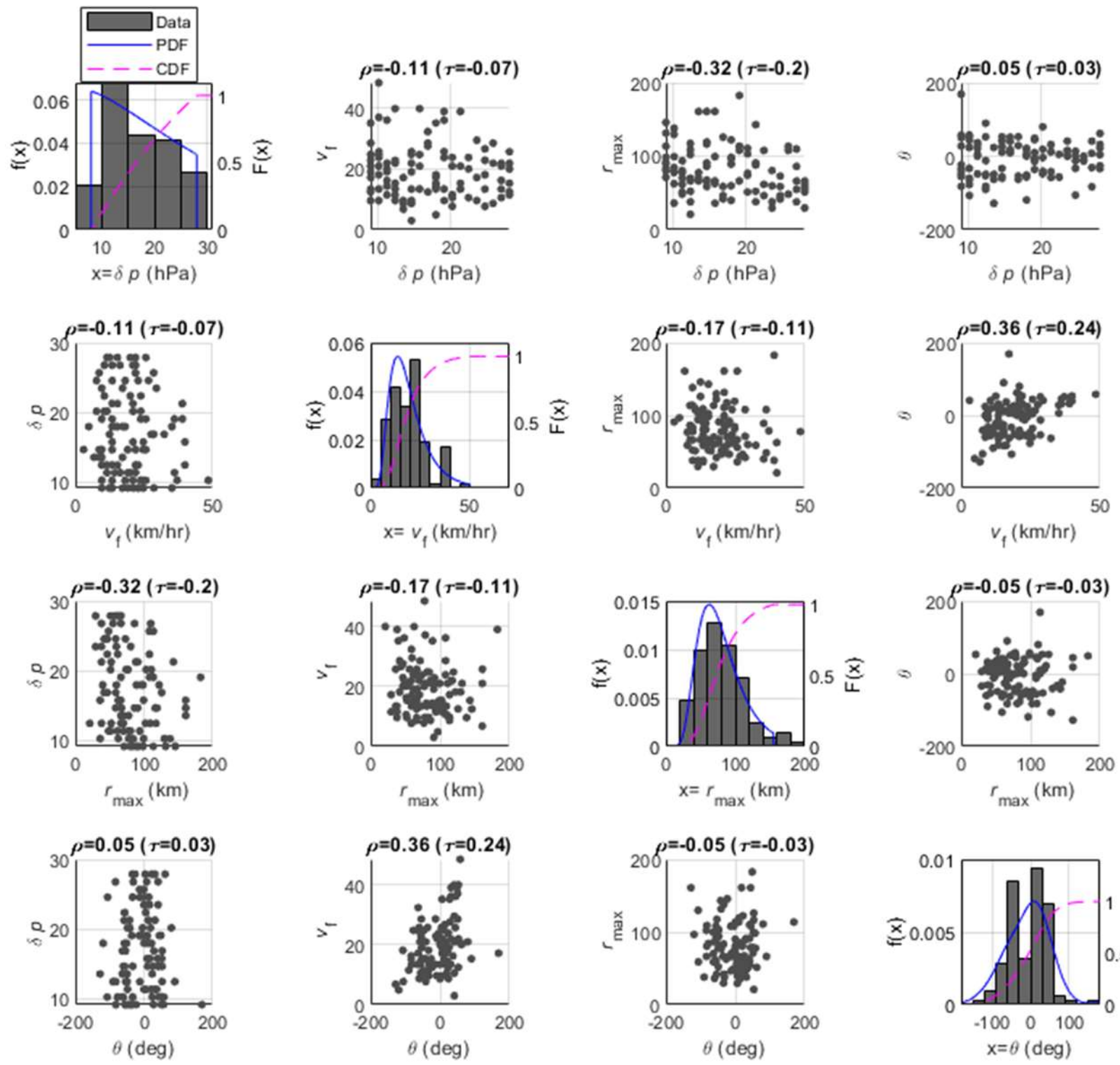

**Fig. 15 Correlation information of TC atmospheric parameters for LI TCs**

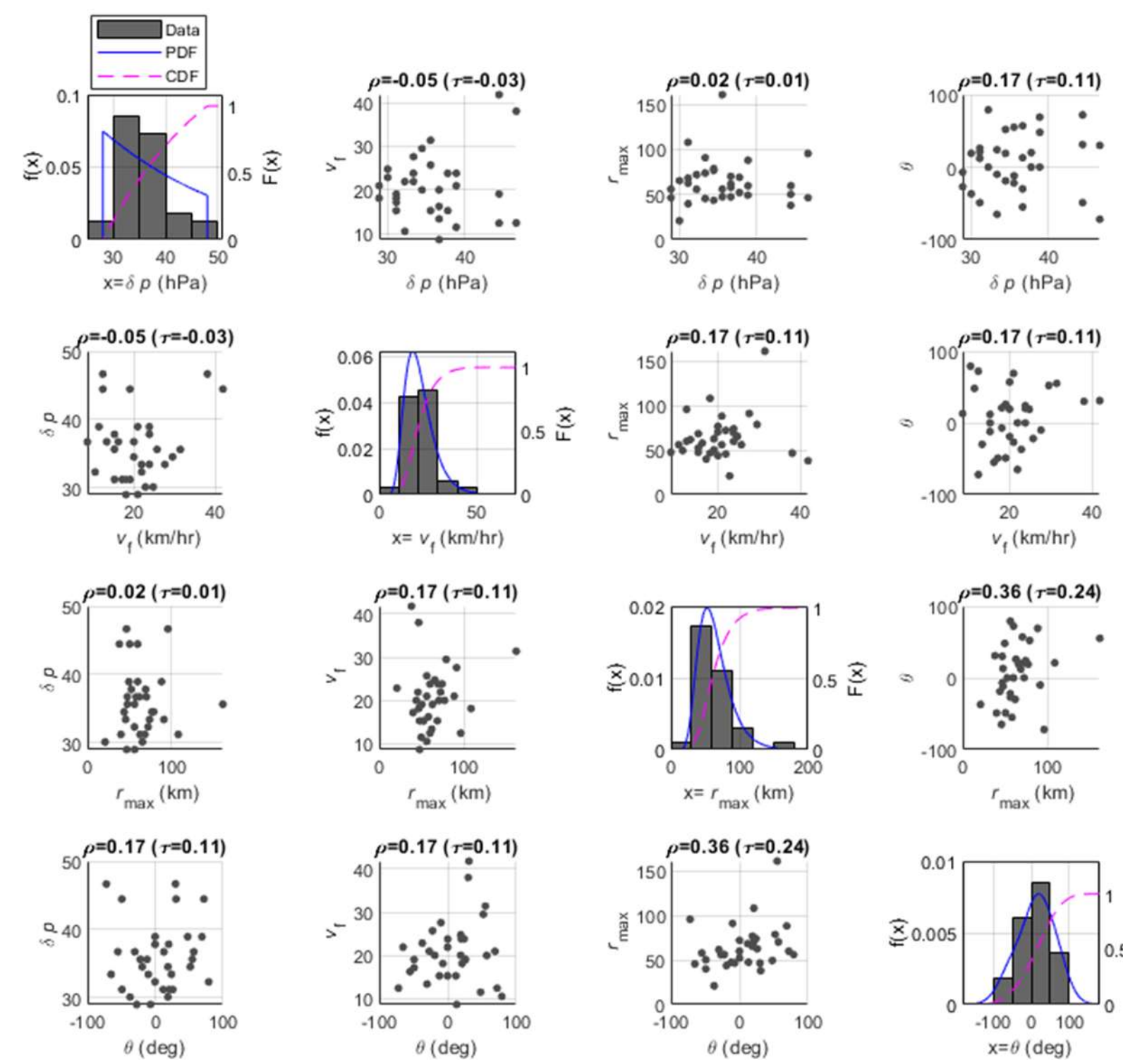

**Fig. 16 Correlation information of TC atmospheric parameters for MI TCs**

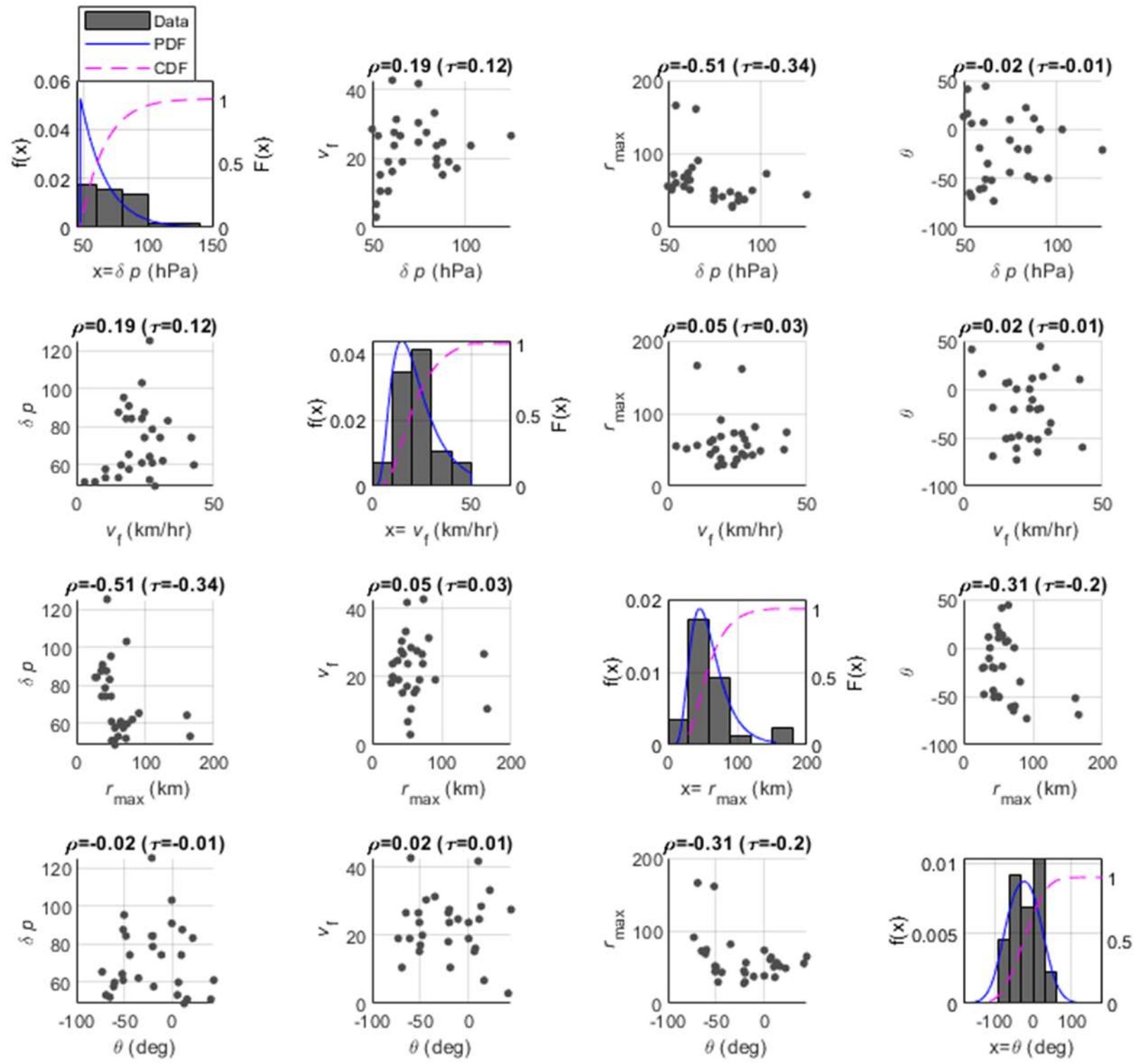

**Fig. 17 Correlation information of TC atmospheric parameters for HI TCs**

**Table 3: The Variable Discretization information of BN nodes**

| NO. | Variable | Discretized values |
|---|---|---|
| 1 | $\Delta P$ (hPa) | 8, 18, 28, 38, 48, 58, 68, 78, 88, 98, 108, 118, 128, 138 |
| 2 | $V_{\mathrm{f}}$ (km/hr) | 5, 10, 15, 20, 25, 30, 35, 40, 45 |
| 3 | $R_{\max}$ (km) | 10, 26.11, 42.22, 58.33, 74.44, 90.56, 106.67, 122.78, 138.89 |
| 4 | $\hat{R}_1$ ($\hat{H}$,m) | Site MR: -0.5, -0.225, 0.05, …, 10.225 (40 bins with equivalent intervals) |
| | | Site GC: -0.1, 0.065, 0.23, …, 6.335 (40 bins with equivalent intervals) |
| 5 | $R_1$ ($H$,m) | Site MR: -0.5, -0.3625, 0.225, …, 10.3625 (80 bins with equivalent intervals) |
| 6 | $\hat{R}_2$ ($\hat{P}$,mm) | Site GC: -0.1,-0.0175, 0.065…, 6.4175 (80 bins with equivalent intervals) |
| 7 | $R_2$ ($P$,mm) | Site MR and Site GC: 0,3.125, 6.25, …, 246.875 (80 bins with equivalent intervals) |
| 8 | $\varepsilon_1$ | -∞,-3,-2,-1,0,1,2,3 |
| 9 | $\varepsilon_2$ | -∞,-3,-2,-1,0,1,2,3 |
| 10 | $\Theta$ (deg) | -80, -60, -40, -20, 0, 20, 40, 60 |
| 11 | $X_0$ | Projected landfall locations as shown in Fig. 18 |

Note: The discretized values presented in rows 1 to 9 provide the lower edge (value) of each bin. In row 10, eight values of Θ are used for TC tracks.

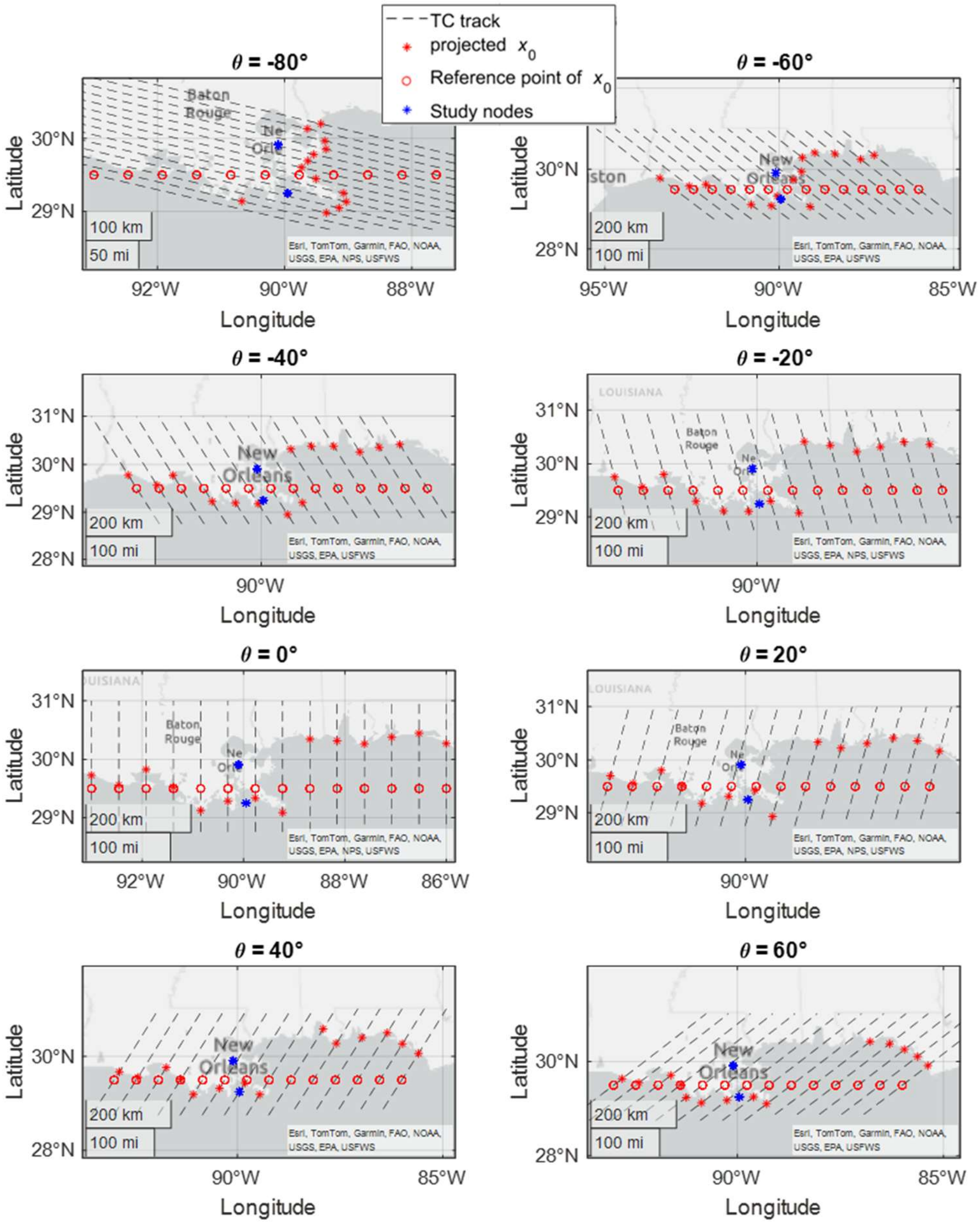

**Fig. 18 Projected landfall locations**

# References

Agrawal, S., N. Hundia, Z. Liu, and M. Bensi. 2025. "Deep learning motivated data imputation of tropical cyclone radius of maximum winds." *14th International Conference on Structural Safety and Reliability - ICOSSAR'25*, Applications of probability, statistics, and machine learning to the diverse field of Catastrophe Modeling. https://doi.org/10.23967/icossar.2025.065.

Al Kajbaf, A., and M. Bensi. 2020. "Application of surrogate models in estimation of storm surge:A comparative assessment." *Applied Soft Computing*, 91: 106184. https://doi.org/10.1016/j.asoc.2020.106184.

Baker, J. W. 2013. "Introduction to Probabilistic Seismic Hazard Analysis."

BayesFusion. 2023. "GeNIe Modeler USER MANUAL."

Bazzurro, P., and C. Allin Cornell. 1999. “Disaggregation of seismic hazard.” *Bulletin of the Seismological Society of America*, 89 (2): 501–520. https://doi.org/10.1785/BSSA0890020501.

Callens, A., D. Morichon, and B. Liquet. 2023. “Bayesian networks to predict storm impact using data from both monitoring networks and statistical learning methods.” *Nat Hazards*, 115 (3): 2031–2050. https://doi.org/10.1007/s11069-022-05625-z.

Chouinard, L. E., and C. Liu. 1997. “Model for Recurrence Rate of Hurricanes in Gulf of Mexico.” *Journal of Waterway, Port, Coastal, and Ocean Engineering*, 123 (3): 113–119. https://doi.org/10.1061/(ASCE)0733-950X(1997)123:3(113).

Couasnon, A., A. Sebastian, and O. Morales-Nápoles. 2018. “A Copula-Based Bayesian Network for Modeling Compound Flood Hazard from Riverine and Coastal Interactions at the Catchment Scale: An Application to the Houston Ship Channel, Texas.” *Water*, 10 (9): 1190. Multidisciplinary Digital Publishing Institute. https://doi.org/10.3390/w10091190.

Dawid, A. P. 1992. “Applications of a general propagation algorithm for probabilistic expert systems.” *Stat Comput*, 2 (1): 25–36. https://doi.org/10.1007/BF01890546.

Demuth, J. L., M. DeMaria, and J. A. Knaff. 2006. “Improvement of Advanced Microwave Sounding Unit Tropical Cyclone Intensity and Size Estimation Algorithms.” *Journal of Applied Meteorology and Climatology*, 45 (11): 1573–1581. American Meteorological Society. https://doi.org/10.1175/JAM2429.1.

Durap, A. 2024. “Mapping coastal resilience: a Gis-based Bayesian network approach to coastal hazard identification for Queensland’s dynamic shorelines.” *Anthropocene Coasts*, 7 (1): 23. https://doi.org/10.1007/s44218-024-00060-y.

Fang, H.-B., K.-T. Fang, and S. Kotz. 2002. “The Meta-elliptical Distributions with Given Marginals.” *Journal of Multivariate Analysis*, 82 (1): 1–16. https://doi.org/10.1006/jmva.2001.2017.

Flores, M. J., R. F. Ropero, and R. Rumí. 2019. “Assessment of flood risk in Mediterranean catchments: an approach based on Bayesian networks.” *Stoch Environ Res Risk Assess*, 33 (11): 1991–2005. https://doi.org/10.1007/s00477-019-01746-3.

Garzon, J. L., O. Ferreira, T. A. Plomaritis, A. C. Zózimo, C. J. E. M. Fortes, and L. V. Pinheiro. 2024. “Development of a Bayesian network-based early warning system for storm-driven coastal erosion.” *Coastal Engineering*, 189: 104460. https://doi.org/10.1016/j.coastaleng.2024.104460.

Gharehtoragh, M. A., and D. R. Johnson. 2024. “Using surrogate modeling to predict storm surge on evolving landscapes under climate change.” *npj Nat. Hazards*, 1 (1): 1–9. Nature Publishing Group. https://doi.org/10.1038/s44304-024-00032-9.

Ho, F. P., and V. A. Myers. 1975. *Joint Probability Method of Tide Frequency Analysis Applied to Apalachicola Bay and St. George Sound, Florida*.

Jensen, F. 1990. “Bayesian updating in recursive graphical models by local comutations.” *Computational Statistics and Data Analysis*, 4: 269–282.

Jia, G., and A. A. Taflanidis. 2013. “Kriging metamodeling for approximation of high-dimensional wave and surge responses in real-time storm/hurricane risk assessment.” *Computer Methods in Applied Mechanics and Engineering*, 261: 24–38. Elsevier.

Landsea, C. W., and J. L. Franklin. 2013. “Atlantic Hurricane Database Uncertainty and Presentation of a New Database Format.” *Monthly Weather Review*, 141 (10): 3576–3592. American Meteorological Society. https://doi.org/10.1175/MWR-D-12-00254.1.

Lauritzen, S. L., and D. J. Spiegelhalter. 1988. “Local Computations with Probabilities on Graphical Structures and Their Application to Expert Systems.” *Journal of the Royal Statistical Society: Series B (Methodological)*, 50 (2): 157–194. https://doi.org/10.1111/j.2517-6161.1988.tb01721.x.

Liu, Z., A. Al Kajbaf, and M. Bensi. 2024a. “Applications of Statistical Learning Methods in Natural Hazard Assessment.” Sendai, Japan.

Liu, Z., M. L. Carr, N. C. Nadal-Caraballo, L. A. Aucoin, M. C. Yawn, and M. Bensi. 2024b. “Comparative Analysis of Joint Distribution Models for Tropical Cyclone Atmospheric Parameters in Probabilistic Coastal Hazard Analysis.” *Stochastic Environmental Research and Risk Assessment*, 38: 1741–1767. https://doi.org/10.1007/s00477-023-02652-5.

Liu, Z., M. L. Carr, N. C. Nadal-Caraballo, L. A. Aucoin, M. C. Yawn, and M. T. Bensi. 2025. “A Multi-Tiered Bayesian Network Coastal Compound Flood Analysis Framework.” *Coastal Engineering*, 104895. https://doi.org/10.1016/j.coastaleng.2025.104895.

Liu, Z., M. L. Carr, N. C. Nadal-Caraballo, M. C. Yawn, A. A. Taflanidis, and M. Bensi. 2024c. "Machine Learning Motivated Data Imputation of Storm Data Used in Coastal Hazard Assessments." *Coastal Engineering*, 190. https://doi.org/10.1016/j.coastaleng.2024.104505.

Lu, P., N. Lin, K. Emanuel, D. Chavas, and J. Smith. 2018. "Assessing Hurricane Rainfall Mechanisms Using a Physics-Based Model: Hurricanes Isabel (2003) and Irene (2011)." *Journal of the Atmospheric Sciences*, 75 (7): 2337–2358. American Meteorological Society. https://doi.org/10.1175/JAS-D-17-0264.1.

McGuire, R. K. 1995. "Probabilistic seismic hazard analysis and design earthquakes: Closing the loop." *Bulletin of the Seismological Society of America*, 85 (5): 1275–1284. https://doi.org/10.1785/BSSA0850051275.

Mohammadi, S., M. T. Bensi, S.-C. Kao, and S. T. DeNeale. 2021. *Multi-Mechanism Flood Hazard Assessment: Example Use Case Studies*. Oak Ridge National Lab.

Mohammadi, S., M. T. Bensi, S.-C. Kao, S. T. DeNeale, J. Kanney, E. Yegorova, and M. L. Carr. 2023. "Bayesian-Motivated Probabilistic Model of Hurricane-Induced Multimechanism Flood Hazards." *Journal of Waterway, Port, Coastal, and Ocean Engineering*, 149 (4): 04023007. American Society of Civil Engineers. https://doi.org/10.1061/JWPED5.WWENG-1921.

Myers, V. A. 1954. *Characteristics of United States Hurricanes Pertinent to Levee Design for Lake Okeechobee, Florida*. U.S. Government Printing Office.

Myers, V. A. 1970. "Joint Probability Method of Tide Frequency Analysis Applied to Atlantic City and Long Beach Island."

Nadal-Caraballo, N. C., M. O. Campbell, V. M. Gonzalez, M. J. Torres, J. A. Melby, and A. A. Taflanidis. 2020. "Coastal Hazards System: A Probabilistic Coastal Hazard Analysis Framework." *Journal of Coastal Research*, 95 (SI): 1211–1216. https://doi.org/10.2112/SI95-235.1.

Nadal-Caraballo, N. C., J. A. Melby, V. M. Gonzalez, and A. T. Cox. 2015. *North Atlantic Coast Comprehensive Study–Coastal Storm Hazards from Virginia to Maine*. Vicksburg, MS: US Army Engineer Research and Development Center.

Nadal-Caraballo, N. C., M. C. Yawn, L. A. Aucoin, M. L. Carr, J. A. Melby, E. Ramos-Santiago, F. A. Garcia-Moreno, V. M. Gonzalez, T. C. Massey, M. B. Owensby, A. A. Taflanidis, A. P. Kyprioti, A. T. Cox, and J. Gonzalez-Lopez. 2022a. *Coastal Hazards System – Puerto Rico (CHS-PR)*. Vicksburg, MS: US Army Engineer Research and Development Center.

Nadal-Caraballo, N. C., M. C. Yawn, L. A. Aucoin, M. L. Carr, J. A. Melby, E. Ramos-Santiago, V. M. Gonzalez, A. A. Taflanidis, A. A. Kyprioti, Z. Cobell, and A. T. Cox. 2022b. *Coastal Hazards System–Louisiana (CHS-LA)*. Vicksburg, MS: US Army Engineer Research and Development Center.

NOAA. 2022. "Re-Analysis Project Hurricane Database." https://www.aoml.noaa.gov/hrd/hurdat/Data_Storm.html.

RAMMB/CIRA. 2021. "The Tropical Cyclone Extended Best Track Dataset (EBTRK)." *Regional and Mesoscale Meteorology Branch*. Accessed March 28, 2022. https:// rammb2. cira. colos tate. edu/ resea rch/ tropi cal-cyclo nes/ tc_ exten ded_ best_ track_ datas et/.

Russell, L. R. 1969. *Probability distributions for Texas Gulf coast hurricane effects of engineering interest.* Stanford University.

Salgado, K., M. L. Martínez, O. Pérez-Maqueo, M. Equihua, I. Mariño-Tapia, and P. Hesp. 2024. "Estimating storm-related coastal risk in Mexico using Bayesian networks and the occurrence of natural ecosystems." *Nat Hazards*, 120 (6): 5919–5940. https://doi.org/10.1007/s11069-024-06460-0.

Sanuy, M., J. A. Jiménez, and N. Plant. 2020. "A Bayesian Network methodology for coastal hazard assessments on a regional scale: The BN-CRAF." *Coastal Engineering*, 157: 103627. https://doi.org/10.1016/j.coastaleng.2019.103627.

Sebastian, A., E. J. C. Dupuits, and O. Morales-Nápoles. 2017. "Applying a Bayesian network based on Gaussian copulas to model the hydraulic boundary conditions for hurricane flood risk analysis in a coastal watershed." *Coastal Engineering*, 125: 42–50. https://doi.org/10.1016/j.coastaleng.2017.03.008.

Taylor, C. C. 2008. "Automatic bandwidth selection for circular density estimation." *Computational Statistics & Data Analysis*, 52 (7): 3493–3500. Elsevier.

Toro, G. R. 2008. *Joint Probability Analysis of Hurricane Flood Hazards for Mississippi, Final report in support of the FEMA-HMTAP flood study of the State of Mississippi (Rev. 1).*

Toro, G. R., A. W. Niedoroda, C. W. Reed, and D. Divoky. 2010. "Quadrature-based approach for the efficient evaluation of surge hazard." *Ocean Engineering*, A Forensic Analysis of Hurricane Katrina's Impact: Methods and Findings, 37 (1): 114–124. https://doi.org/10.1016/j.oceaneng.2009.09.005.

Xi, D., N. Lin, and J. Smith. 2020. "Evaluation of a Physics-Based Tropical Cyclone Rainfall Model for Risk Assessment." *Journal of Hydrometeorology*, 21 (9): 2197–2218. American Meteorological Society. https://doi.org/10.1175/JHM-D-20-0035.1.

Zhu, L., S. M. Quiring, and K. A. Emanuel. 2013. "Estimating tropical cyclone precipitation risk in Texas." *Geophysical Research Letters*, 40 (23): 6225–6230. https://doi.org/10.1002/2013GL058284.